\documentclass[a4paper,fleqn]{cas-sc}
\usepackage{subcaption}
\usepackage[authoryear]{natbib}

\def\tsc#1{\csdef{#1}{\textsc{\lowercase{#1}}\xspace}}

\def\pder#1#2{\frac{\partial #1}{\partial #2}}
\def\tder#1#2{\frac{\mathrm{d} #1}{\mathrm{d} #2}}
\def\ltder#1#2{\mathrm{d}#1/\mathrm{d} #2}
\def\mtder#1#2#3{\frac{{\mathrm{d}^{#1}#2}}{\mathrm{d} #3^{#1}}}
\def\lmtder#1#2#3{\mathrm{d}^{#1}#2/\mathrm{d} #3^{#1}}

\def\vect#1{\mathbf{#1}}
\def\vecg#1{\boldsymbol{#1}}

\def\D#1{~\mathrm{d}#1}
\tsc{WGM}
\tsc{QE}
\tsc{EP}
\tsc{PMS}
\tsc{BEC}
\tsc{DE}

\def\SI#1#2{\ensuremath{#1\hspace{1.5pt}\text{#2}}}
\def\unexp#1#2{\ensuremath{\text{#1}^{#2}}}
\def\GPa{GPa}

\def\GHz{GHz}
\def\kg{kg}
\def\nN{nN}

\def\m{m}
\def\nm{nm}

\newdefinition{rmk}{Remark}

\usepackage[demo]{adjustbox}
\usepackage{xcolor}
\usepackage{atbegshi}
\AtBeginDocument{\AtBeginShipoutNext{\AtBeginShipoutDiscard}}	
\PassOptionsToPackage{hyphens}{url}\usepackage{hyperref} 

\begin{document}
	\pagestyle{empty} 
	\begin{titlepage}
		\color[rgb]{.4,.4,1}
		\hspace{5mm}

		\bigskip
		
		\hspace{15mm}
		\begin{minipage}{10mm}
			\color[rgb]{.7,.7,1}
			\rule{1pt}{226mm}
		\end{minipage}
		\begin{minipage}{133mm}
			\vspace{10mm}        
			\color{black}
			\sffamily
			\LARGE\bfseries On the dynamics of nano-frames  \\[-0.3\baselineskip]   \\[-0.3\baselineskip] 
			
			\vspace{5mm}
			{\large {Preprint of the article published in \\[-0.4\baselineskip] International Journal of Engineering Science \\[-0.1\baselineskip] 160, March 2021, 103433 }} 
			
			\vspace{10mm}        
			{\large Andrea Francesco Russillo,\\[-0.4\baselineskip] \textsc{Giuseppe Failla}, \\[-0.4\baselineskip] Gioacchino Alotta,
			\\[-0.4\baselineskip] Francesco Marotti de Sciarra,
			\\[-0.4\baselineskip] Raffaele Barretta} 
			
			\large
			
			\vspace{40mm}
			\vspace{5mm}
			
			\small 
			\url{https://doi.org/10.1016/j.ijengsci.2020.103433}
			
			\textcircled{c} 2021. This manuscript version is made available under the CC-BY-NC-ND 4.0 license \url{http://creativecommons.org/licenses/by-nc-nd/4.0/}
			\hspace{30mm} 
			\color[rgb]{.4,.4,1} 
		\end{minipage}
	\end{titlepage}
	
\let\WriteBookmarks\relax
\def\floatpagepagefraction{1}
\def\textpagefraction{.001}
\shorttitle{On the dynamics of nano-frames}
\shortauthors{A.F. Russillo, G. Failla, G. Alotta, F. Marotti de Sciarra, R. Barretta}
                  
\title [mode = title]{On the dynamics of nano-frames} 

\author[1]{Andrea Francesco Russillo}
\author[1]{Giuseppe Failla}
\author[1]{Gioacchino Alotta}
\author[2]{Francesco {Marotti de Sciarra}}
\author[2]{Raffaele Barretta}[orcid=https://orcid.org/0000-0002-8535-0581]
\cormark[1]
\ead[1]{rabarret@unina.it}
\cortext[cor1]{Corresponding author}

\address[1]{Department of Civil, Environmental, Energy and Materials Engineering (DICEAM), University of Reggio Calabria, Via Graziella, 89124 Reggio Calabria, Italy}

\address[2]{Department of Structures for Engineering and Architecture, University of Naples Federico II, via Claudio 21, 80125 Naples, Italy}

\begin{abstract}
In this paper, size-dependent dynamic responses of small-size frames are modelled by stress-driven nonlocal elasticity and assessed by a consistent finite-element methodology. 
Starting from uncoupled axial and bending differential equations, the exact dynamic stiffness matrix of a two-node stress-driven nonlocal beam element is evaluated in a closed form. 
The relevant global dynamic stiffness matrix of an arbitrarily-shaped small-size frame, where every member is made of a single element, is built by a standard finite-element assembly procedure. 
The Wittrick-Williams algorithm is applied to calculate natural frequencies and modes.
The developed methodology, exploiting the one conceived for straight beams in [International Journal of Engineering Science 115, 14-27 (2017)], is suitable for investigating size-dependent free vibrations of small-size systems of current applicative interest in Nano-Engineering, such as carbon nanotube networks and polymer-metal micro-trusses.
\end{abstract}

\maketitle

\section{Introduction}
\begin{figure}
\centering
\begin{subfigure}[pos=h]{0.45\linewidth}
\centering
\includegraphics[scale=0.15]{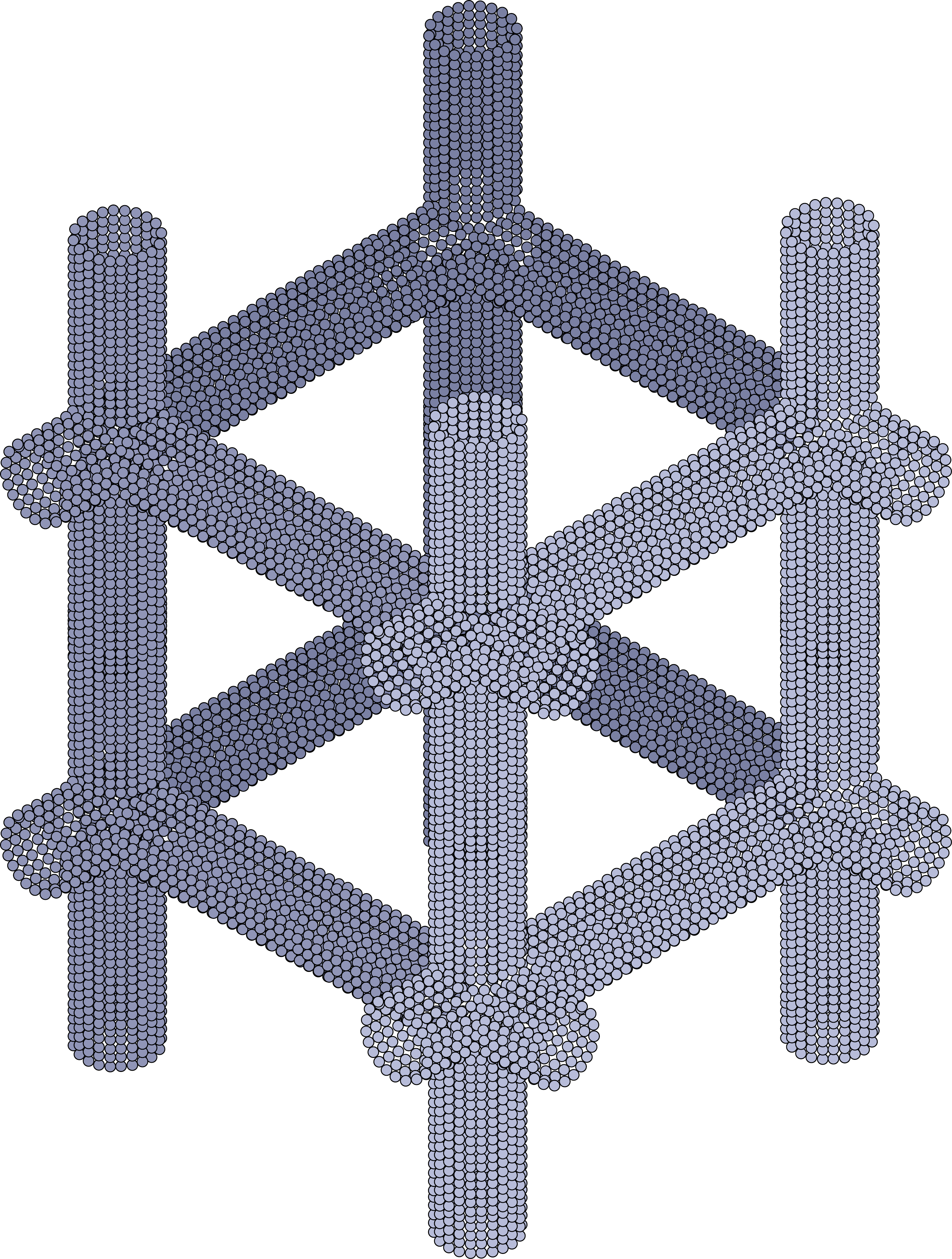}
\caption{}
\end{subfigure}
\begin{subfigure}[pos=h]{0.45\linewidth}
\centering
\vspace{1.46cm}
\includegraphics[scale=0.25]{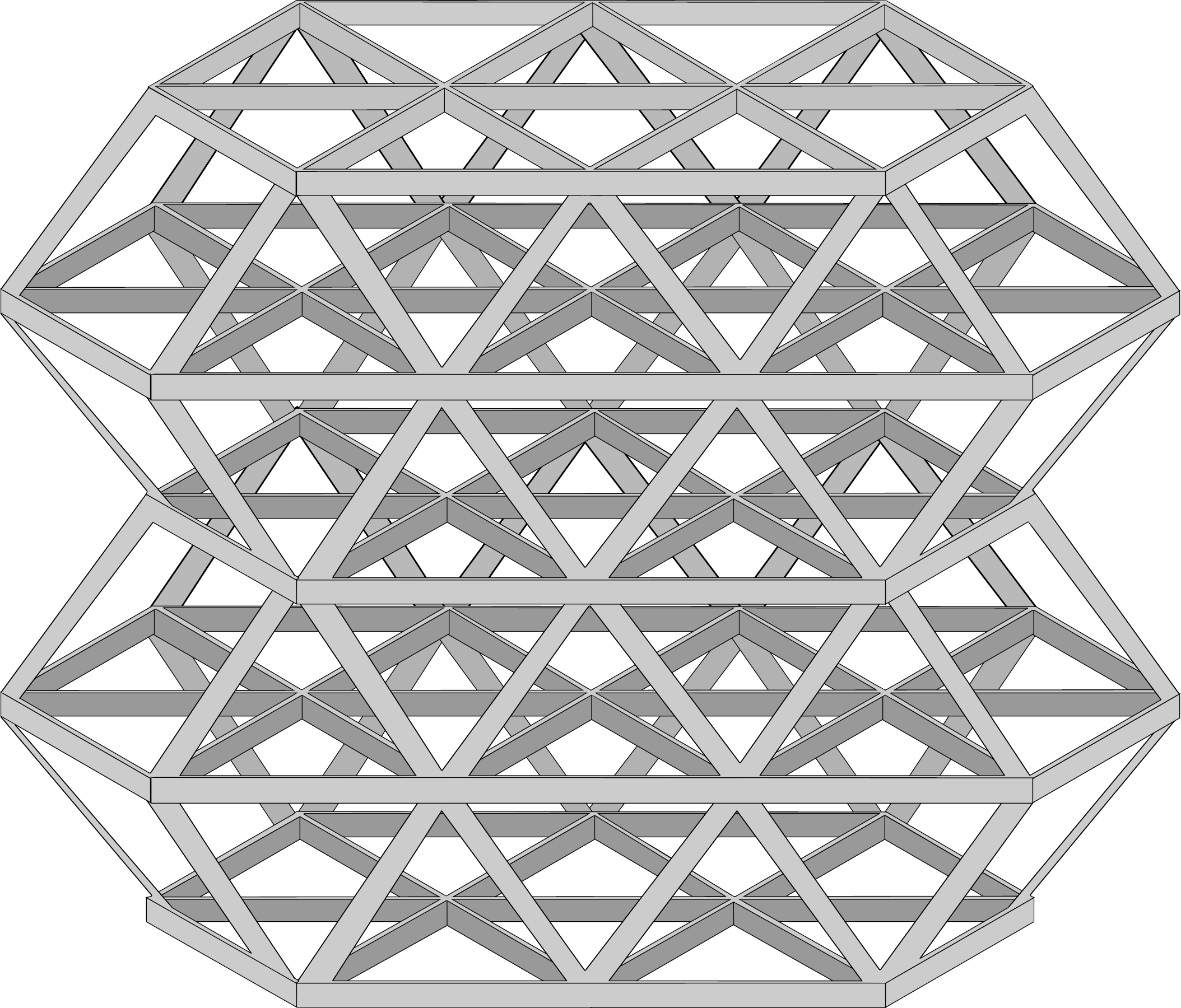}
\caption{}
\end{subfigure}
\caption{\label{fig:nanostruct}Typical advanced materials and structures in Engineering Science: (a) carbon nanotube network \citep{ZHANG201838}, (b) polymer-metal micro-truss \citep{JUAREZ2018442}.}
\end{figure}

Small-size structures as carbon nanotube networks \citep{ZHANG201838}, 3D-printed polymer-metal micro-trusses \citep{JUAREZ2018442} and ceramic nanolattices \citep{Meza1322} are attracting a considerable interest for remarkable features not obtainable by standard materials. Properties as high-thermal conductivity, excellent mechanical strength, electrical conductivity, high-strain sensitivity and large surface area make carbon nanotube networks 
\citep{ZHANG201838}
ideally suitable for the next generation of thermal management \citep{FASANO20151028}
and electronic nanodevices \citep{Lee2016}, 
strain sensors \citep{CHAO2020105187} 
and hydrogen storage \citep{OZTURK2015403,BI202017637}. 
Polymer-metal micro-trusses exhibit enhanced strength, conductivity and electrochemical properties, while ceramic nanolattices feature highest strength- and stiffness-to-weight ratios \citep{ZhangSmall}. 
In view of promising applications in a large number of fields of Engineering Science, great attention is currently devoted to small-size structures
\citep{GHAYESH20198};  for an insight, typical geometries currently under investigation are shown in Figure \ref{fig:nanostruct}.\\

There exist accurate yet computationally very demanding mechanical models of small-size structures, e.g. those involving molecular dynamics simulation for carbon nanotube networks \citep{BARRETTA20171,GENOESE2017316}. 
On the other hand, several studies have focused on developing analytical or numerical models of small-size continua, which may provide rigorous insight into the essential mechanics of the system and be readily implementable for design and optimization at a relatively-low computational effort. For this purpose, a typical approach is the formulation of continua enriched with nonlocal terms capable of capturing size effects that, instead, cannot be described by the free-scale local continuum approach. Now, nonlocal theories represent a rather established approach to investigate small-size continua. Among others, typical examples are Eringen's integral theory \citep{Eringen1972,Eringen19834703}, strain-gradient theories \citep{Aifantis1999189,aifantis2003update,aifantis2009exploring,aifantis2011gradient,askes2011gradient,
challamel2016nonlocal,polizzotto2014stress,polizzotto2015unifying}, micropolar ``Cosserat'' theory \citep{lakes1991experimental}, peridynamic theory \citep{SILLING2000175,SILLING2007} and
mechanically-based approaches involving long-range interactions among non-adjacent volumes \citep{PAOLA20102347}. 
Surveys of progress regarding nonlocal elasticity and generalized continua
can be found in \citep{RomanoMeccanica2020} and \citep{RomanoMicro}, respectively. \\

\break

Most of the existing nonlocal theories have developed nonlocal models of $1$D and $2$D structures, whose statics and dynamics have been investigated under various boundary conditions in a considerable number of studies, such as: \citep{akgoz2013size,ATTIA20181,challamel2018static,DASTJERDI2019125,%
dipaola2009,di2013non,FARAJPOUR2018231,fuschi2019size,GHAYESH2019103139,%
GHOLIPOUR2020103221,KARAMI2020103309,KHANIKI201923,%
LI201881,LI2020103311,MALIKAN2020103210,%
numanouglu2018dynamic,pinnola2020variationally,%
SHE201958,SRIVIDHYA20181,ZHANG2020103317}. 
In this context, an effective approach is the so-called stress-driven nonlocal model, relying on the idea that elastic deformation fields are output of convolution integrals between stress fields and appropriate averaging kernel \citep{ROMANO201714}.
Nonlocal integral convolution, endowed with the special bi-exponential kernel, can be conveniently replaced with a higher-order differential equation supplemented with non-standard constitutive boundary conditions.
The stress-driven theory leads to well-posed structural problems \citep{ROMANO2017184}, does not exhibit paradoxical results typical of alternative nonlocal beam models \citep{Challamel2008,
demir2017analysis,fernandez2016bending} and, in the last few years, has gained increasing popularity for consistency, robustness and ease of implementation. 
Stress-driven nonlocal theory of elasticity has been applied to several problems of nanomechanics, as witnessed by recent contributions regarding buckling \citep{OskouieIJCMSE,DARBAN2020103338}, bending \citep{OskouieActaSin,OskouieEPJP,ZHANG2020112362,Roghani2020}, axial \citep{barretta2019longitudinal} and torsional responses \citep{barretta2018stress} of nano-beams and elastostatic behaviour of nano-plates \citep{BARRETTA201938,FARAJPOUR2020103368}.

Nonlocal finite-element formulations have been proposed, in general to discretize single beams or rods \citep{de2014finite,aria2019nonlocal,alotta2014finite,Alotta2017,alotta2017finite}. 
A very recent study, however, has posed the issue of addressing the dynamics of small-size 2D frames/trusses \citep{NUMANOGLU2019103164}, made by assembling nonlocal beams/rods. 
Eringen's differential law has been adopted and the principle of virtual work has been used to derive separate stiffness and mass matrices of a two-node nonlocal element. 
Typical shape functions of a two-node local element have been exploited, with bending modelled by Bernoulli-Euler kinematic theory. 
A further recent contribution in this field has been given by \citet{hozhabrossadati2020free}, who developed a two-node six-degree-of-freedom beam element for free-vibrations of 3D nano-grids. 
Separate stiffness and mass matrices have been derived treating axial, bending and torsional responses by Eringen's differential law and weighted residual method. 

However, the studies by \cite{NUMANOGLU2019103164} and \cite{hozhabrossadati2020free}, 
dealing with dynamics of nanostructural systems via nonlocal finite elements,
are based on Eringen's differential formulation which leads to mechanical paradoxes and unviable elastic responses \citep{peddieson2003application}, 
a conclusion acknowledged by the community of Engineering Science \citep{fernandez2016bending}.
Lack of alternative technically significant contributions on the matter may also be attributed to the fact that not all size-dependent theories allow for formulating stiffness, mass matrices and nodal forces to be assembled in $2$D and $3$D frames/trusses. 
As a matter of fact, there is a great interest in developing accurate and computationally-effective nonlocal models of complex nanostructures in view of their increasing relevance in several engineering fields: carbon nanotube networks \citep{ZHANG201838}, polymer-metal micro-trusses \citep{JUAREZ2018442}, ceramic nanolattices \citep{Meza1322}.
Various examples of nano/micro-scale hierarchical lattice structures and cellular nanostructures have been pointed out by \cite{NUMANOGLU2019103164} along with several related applications.\\

This paper proposes an effective approach to model and assess the dynamic behaviour of complex small-size frames, exploiting the treatment by \citet{ROMANO201714} confined to straight nano-beams. 
Key novelties are as follows.
\begin{enumerate}
\item
Adoption of a well-posed and experimentally consistent stress-driven nonlocal formulation to capture size effects within the members of the structure, assuming uncoupled axial and bending motions (small displacements).
\item
Derivation of the exact dynamic stiffness matrix of a two-node stress-driven nonlocal element, from which the global dynamic stiffness matrix of the structure can be readily built by a standard finite-element assembly procedure.
 \end{enumerate}
Upon constructing the global dynamic stiffness matrix, all natural frequencies and related modes of the structure are calculated using the Wittrick-Williams (WW) algorithm. The formulation applies not only to frames, but also to trusses. It is presented for 2D frames and is readily extendable to 3D networks of nanotechnological interest. 

The main advantages of the proposed approach are summarized as follows. The stress-driven methodology is not affected by inconsistencies and paradoxes corresponding to alternative nonlocal models \citep{ROMANO2017151}. 
The dynamic-stiffness approach captures the exact dynamic response, using a single two-node beam element for every frame member without any internal mesh. Further, the WW algorithm provides all natural frequencies exactly, without missing anyone and including multiple ones.

The paper is organized as follows. 
The stress-driven nonlocal formulation for axial and bending motions is described in Section~\ref{sec:2}. 
The exact dynamic stiffness matrix of two-node stress-driven nonlocal truss and beam elements is established in Section~\ref{sec:3}. 
In addition, the assembly procedure to build the global dynamic stiffness matrix of arbitrarily-shaped small-size frames is illustrated therein. 
The implementation of the WW algorithm is discussed in Section~\ref{sec:4}. 
Numerical applications are presented in Section~\ref{sec:5}, investigating the role of size effects on the free-vibration responses of small-size 2D structures of current technical interest.
\section{Stress-driven nonlocal integral elasticity}\label{sec:2}
This Section introduces fundamental equations governing axial and bending vibrations of a nonlocal beam, according to the stress-driven model recently introduced by \cite{ROMANO201714,ROMANO2017184}. Specifically, axial and bending responses are uncoupled on the assumption of small displacements. 

\begin{figure}[pos=h]
\centering
\includegraphics[width=10cm]{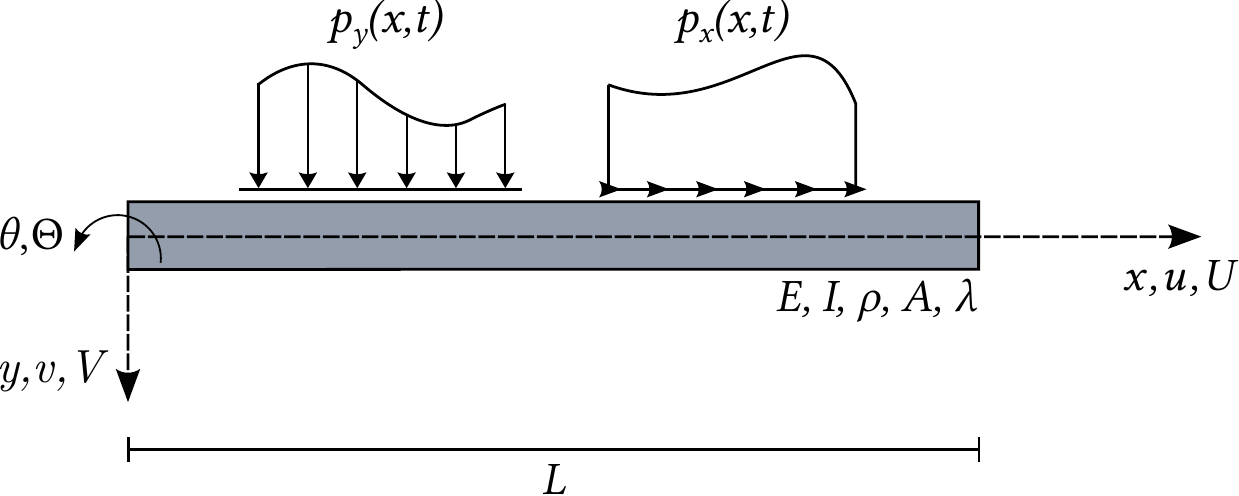}
\caption{Stress-driven nonlocal beam in axial and bending vibrations.}
\label{Fig1}
\end{figure}

Consider a plane beam of length $L$, uniform cross section of area $A$ and moment of inertia $I$, as shown in Fig. \ref{Fig1}. Let $E$ be the local elastic stiffness at the macroscopic scale. 

According to the stress-driven model, the uniaxial strain-stress relationship reads:
\begin{equation}
\varepsilon(x,t)=\frac{1}{E}\int_0^L \Phi(|x-\zeta|)\sigma(\zeta,t)d\zeta
\label{eq2GA}
\end{equation}
where $\varepsilon$ is the elastic longitudinal strain, $\sigma$ is the normal stress, $\Phi(|x-\zeta|)$ is a scalar function known as attenuation function assumed to fullfil positivity, symmetric and limit impulsivity. 

\begin{figure}[pos=h]
\centering
\includegraphics[scale=0.9]{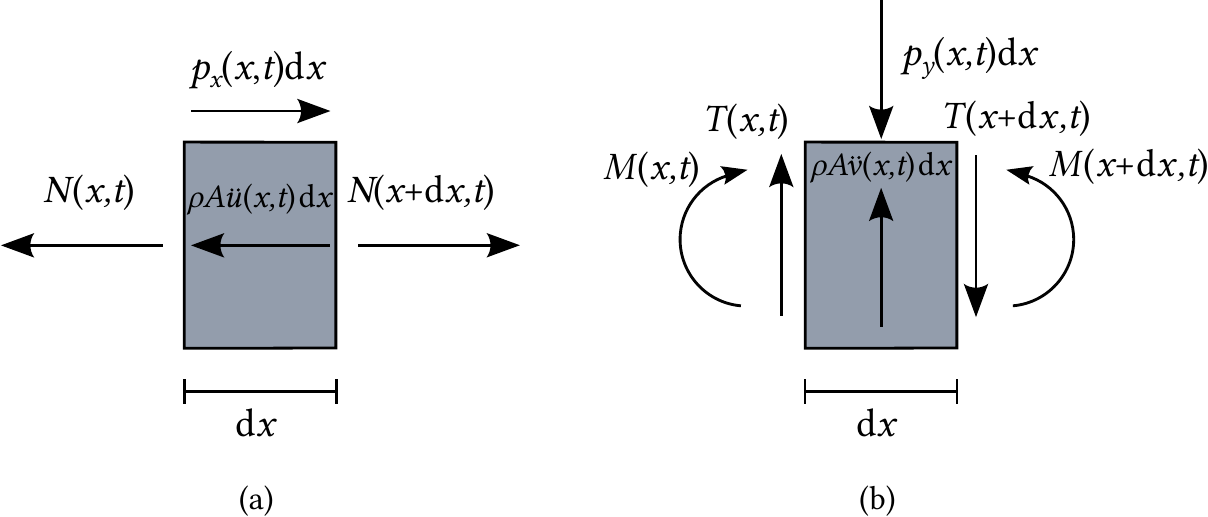}
\caption{Equilibrium of a beam segment: axial problem (a), bending problem (b).}
\label{Fig2}
\end{figure}

First, let us focus on the axial response of the beam in Fig.~\eqref{Fig1}. On the assumption that cross sections remain plane and normal to the longitudinal axis, from Eq.~\eqref{eq2GA} the following equation can be derived between longitudinal generalized strain $\gamma$ and axial force $N$ \citep{barretta2019longitudinal}:
\begin{equation}
\gamma(x,t)=\frac{1}{EA}\int_0^L \Phi(|x-\zeta|)N(\zeta,t)d\zeta
\label{eq10GA}
\end{equation}
being
\begin{equation}\label{eqn:eta}
\gamma(x,t) = u^{(1)}(x,t)
\end{equation}
where $u$ is the axial displacement and superscript $(k)$ means $k^{\mathrm{th}}$ derivative w.r.t. the spatial coordinate $x$. As in recent works \citep{ROMANO201714,ROMANO2017184,barretta2019longitudinal}, $\Phi$ is taken as the following bi-exponential function:
\begin{equation}
\Phi(x,\zeta)=\frac{1}{2L_c}\exp\left(-\frac{|x-\zeta|}{L_c}\right)
\label{eq4GA}
\end{equation}
where $L_c=\lambda\cdot L$ is the characteristic length, being $\lambda$ a material-dependent parameter. Eq.~\eqref{eq4GA} fulfils the requirements of symmetry and positivity. Moreover, Eq.~\eqref{eq4GA} satisfies the property of limit impulsivity, i.e. reverts to the Dirac's delta function for $\lambda \rightarrow 0$, so that Eq.~\eqref{eq10GA} reduces to the standard local constitutive law of linear elasticity at internal points of the structural domain. 
The choice of Eq.~\eqref{eq4GA} for $\Phi$ is motivated by the fact that, using integration by parts, the integral constitutive law \eqref{eq10GA} can be now reverted to the following equivalent differential equation:
\begin{equation}
N(x,t) = - EA\cdot L_{c}^{2}\left(\gamma^{(2)}(x,t) - \frac{1}{L_{c}^2}\gamma(x,t) \right)
\label{eq11GA}
\end{equation}
with the additional constitutive BCs
\begin{equation}\label{eq12GA}
\begin{aligned}
&\gamma^{(1)}(0,t)=\frac{1}{L_c}\gamma(0,t)\\
&\gamma^{(1)}(L,t)=-\frac{1}{L_c}\gamma(L,t)
\end{aligned}\refstepcounter{equation}\tag{\theequation a,b}
\end{equation}
Next, consider the equilibrium equation governing the axial vibration response (see Fig.~\ref{Fig2}a)
\begin{equation}
N^{(1)}(x,t) + p_x(x,t) -\rho A \ddot u(x,t)=0
\label{eq13GA}
\end{equation}
where $\rho$ is the volume mass density of the material, $p_{x}(x,t)$ is the external transversal force per unit length. Combining Eq.~\eqref{eq13GA} with Eq.~\eqref{eq11GA} leads to the following partial differential equation governing axial vibrations of a stress-driven nonlocal beam:
\begin{equation}\label{eq14GA}
-EA\cdot L_{c}^{2}\left(u^{(4)}(x,t) - \frac{1}{L_{c}^2}u^{(2)}(x,t) \right) + p_{x}(x,t) - \rho A \ddot{u}(x,t) = 0
\end{equation}
Eq.~\eqref{eq14GA} is a partial differential equation in the unknown time-dependent axial displacement $u$, to be solved enforcing the two classical static/kinematic BCs and the additional constitutive BCs \eqref{eq12GA} together with the initial conditions. As $\lambda \rightarrow0$, Eq.~\eqref{eq14GA} reverts to the classical partial differential equation governing axial vibrations of the local beam.

Further, for the purposes of this study, it is of interest to formulate the equations governing bending vibrations of the stress-driven nonlocal beam. Assuming the Bernoulli-Euler beam model, Eq. \eqref{eq2GA} leads to the following nonlocal relation between elastic curvature $\chi$ and bending moment interaction $M$ \citep{ROMANO201714,ROMANO2017184}
\begin{equation}
\chi(x,t)=\frac{1}{EI}\int_0^L \Phi(|x-\zeta|)M(\zeta,t)d\zeta
\label{eq3GA}
\end{equation}
In Eq. \eqref{eq3GA}
\begin{equation}\label{eqn:curvature}
\chi(x,t)=-v^{(2)}(x,t), \qquad \theta(x,t) = -v^{(1)}(x,t)\refstepcounter{equation}\tag{\theequation a,b}
\end{equation}
where $v$ is the deflection in the $y$ direction, $\theta$ is the rotation (positive counterclockwise). Again, using the bi-exponential function \eqref{eq4GA} in Eq.~\eqref{eq3GA} and integrating by parts leads to the equivalent differential problem of Eq.~\eqref{eq3GA}:
\begin{equation}
M(x) =- EI\cdot L_c^2 \left(\chi^{(2)}(x) - \frac{1}{L_c^2}\chi(x)\right)
\label{eq5GA}
\end{equation}
with the additional constitutive BCs
\begin{equation}
\begin{aligned}
&\chi^{(1)}(0,t)=\frac{1}{L_c}\chi(0,t)\\
&\chi^{(1)}(L,t)=-\frac{1}{L_c}\chi(L,t)
\end{aligned}\refstepcounter{equation}\tag{\theequation a,b}
\label{eq6GA}
\end{equation}

Next, consider that the bending vibration response of a Bernoulli-Euler beam is governed by the  differential condition of equilibrium
\begin{equation}
M^{(2)}(x,t)+p_{y}(x,t)-\rho A \ddot v(x,t)=0
\label{eq8GA}
\end{equation}
where $p_{y}(x,t)$ is the external transverse force per unit length; further,
\begin{equation}
M^{(1)}(x,t)=T(x,t)
\label{eq7GAa}
\end{equation}
as shown in Fig.~\ref{Fig2}b. Combining Eq.~\eqref{eq5GA} and Eq.~\eqref{eq8GA} leads to the following equation governing bending vibrations of a stress-driven nonlocal beam:
\begin{equation}
EI \cdot L_{c}^2 \left(v^{(6)}(x,t) - \frac{1}{L_{c}^2} v^{(4)}(x,t) \right) + p_{y}(x,t) - \rho A \ddot{v}(x,t) = 0
\label{eq9GA}
\end{equation}
Eq. \eqref{eq9GA} is a partial differential equation in the unknown time-dependent deflection $v$, to be solved enforcing the classical four static/kinematic BCs and the two constitutive BCs \eqref{eq6GA} \citep{ROMANO201714,ROMANO2017184}, together with the initial conditions. Note that, as $\lambda\rightarrow0$, Eq. \eqref{eq9GA} reverts to the classical partial differential equation governing the local Bernoulli-Euler beam. 

Eq.~\eqref{eq14GA} and Eq.~\eqref{eq9GA} are the basis to derive the exact dynamic stiffness of a two-node stress-driven nonlocal beam element, as explained in the next Section.
\section{Exact dynamic stiffness matrix of stress-driven nonlocal beam elements}\label{sec:3}
\begin{figure}[pos=h]
\centering
\includegraphics[width=12cm]{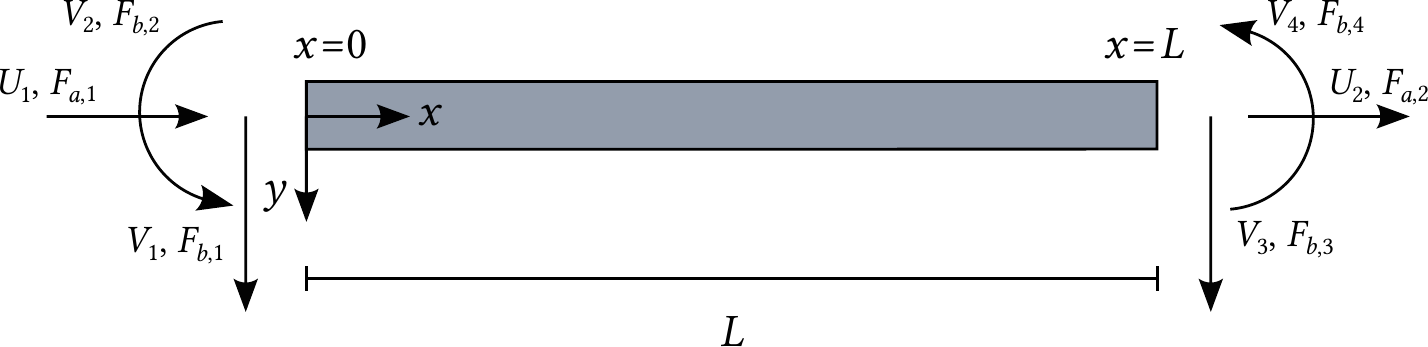}
\caption{Nodal forces and displacements of a two-node stress-driven nonlocal beam element.}
\label{Fig3}
\end{figure}

Let us consider the stress-driven nonlocal beam in Fig.~\ref{Fig3}, acted upon by harmonic forces/moments at the ends.
Beam ends are referred to as ``nodes'' with three degrees of freedom each and the beam as a two-node stress-driven nonlocal beam element. Denoting by $u(x,t) = U(x)\mathrm{e}^{\mathrm{i}\omega t}$, $v(x,t) = V(x)\mathrm{e}^{\mathrm{i}\omega t}$, $\theta(x,t)=\Theta(x)\mathrm{e}^{\mathrm{i}\omega t}$,~\dots, axial and bending response variables, be $\vect{F}\mathrm{e}^{\mathrm{i}\omega t}=[-N(0)\;N(L)\;-T(0)\;-M(0)\;T(L)\;M(L)]^{\mathrm{T}}\mathrm{e}^{\mathrm{i}\omega t}$ the vector of nodal forces and $\vect{U}\mathrm{e}^{\mathrm{i}\omega t}=[U(0)\;U(L)\;V(0)\;\Theta(0)\;V(L)\;\Theta(L)]^{\mathrm{T}}\mathrm{e}^{\mathrm{i}\omega t}$ the corresponding vector of nodal displacements.

For generality, the dynamic stiffness matrix of the two-node stress-driven nonlocal beam element is sought in terms of dimensionless frequencies \citep{banerjee1996,banerjee1998,banerjee2001,banerjee2003}. 
Thus, the equations of motion \eqref{eq14GA} and \eqref{eq9GA} governing steady-state responses under harmonic forces/moments at beam ends and the associated constitutive BCs \eqref{eq12GA} and \eqref{eq6GA} are rewritten as ($\omega$\,--\,dependence of the response variables is omitted for brevity): 
\\
\begin{equation}
(\mathcal{L}_{a}+\overline{\omega}_{a}^{2}\mathcal{I})[U]=0\,,
\qquad\textrm{with}\qquad
\mathcal{L}_{a}
:=
-\lambda^2 \mtder{4}{}{\xi} + \mtder{2}{}{\xi}\,,
\quad
\xi\in[0,1]
\label{eq16GA}
\end{equation}
\begin{equation}
\begin{aligned}
&\left.\tder{\Gamma}{\xi}\right\rvert_{\xi=0}=\frac{1}{\lambda}\Gamma(0)\\
&\left.\tder{\Gamma}{\xi}\right\rvert_{\xi=1} =-\frac{1}{\lambda}\Gamma(1)
\end{aligned}\refstepcounter{equation}\tag{\theequation a,b}
\label{eqn:dimensionless_gen_strain}
\end{equation}
\begin{equation}
(\mathcal{L}_{b}+\overline{\omega}_{b}^{4}\mathcal{I})[V]=0\,,
\qquad\textrm{with}\qquad
\mathcal{L}_{b}
:=
\lambda^2 \mtder{6}{}{\xi} -\mtder{4}{}{\xi} \,,
\quad
\xi\in[0,1]
\label{eq17GA}
\end{equation}
\begin{equation}
\begin{aligned}
&\left.\tder{X}{\xi}\right\rvert_{\xi=0}=\frac{1}{\lambda}X(0)\\
&\left.\tder{X}{\xi}\right\rvert_{\xi=1} =-\frac{1}{\lambda}X(1)
\end{aligned}\refstepcounter{equation}\tag{\theequation a,b}
\label{eqn:dimensionless_curvature}
\end{equation}
where $\,\xi = x/L\,$, $\,\lambda = L_{c}/L\,$,  $\,\overline{\omega}_a = [\rho A (\omega L)^{2}/EA]^{1/2}\,$,  $\,\overline{\omega}_b = [\rho A (\omega L^2)^{2}/EI]^{1/4}\,$ and 
$\,\mathcal{I}\,$ is the identity map. 
Further, in Eq.~\eqref{eq16GA} through Eqs.~\eqref{eqn:dimensionless_curvature}, the notation $\lmtder{k}{}{\xi}$ distinguishes the derivative w.r.t. to the dimensionless spatial coordinate $\xi$ from the derivative w.r.t. to $x$, indicated by the superscript $(k)$.
Upon enforcing the constitutive BCs \eqref{eqn:dimensionless_gen_strain} and \eqref{eqn:dimensionless_curvature}, the exact solutions of Eq.~\eqref{eq16GA} and Eq.~\eqref{eq17GA} can be obtained in the following analytical forms:
\begin{align}
&U(\xi) = \sum_{k=1}^{2}c_{a,k} f_k (\xi) \label{eqn:u} \\
&V(\xi) = \sum_{k=1}^{4}c_{b,k} g_k (\xi) \label{eqn:v}
\end{align}
where $c_{a,k}$ for $k=1,2$ and $c_{b,k}$ for $k=1,...,4$ are integration constants depending on the classical static/kinematic BCs. Further, $f_{k}(\xi)=f_{k}(\xi,\overline{\omega}_a)$ and $g_{k}(\xi)=g_{k}(\xi,\overline{\omega}_b)$  are closed analytical functions depending on frequency and parameters of the stress-driven nonlocal beam, reported in Appendix A for brevity. 

Now, from Eq.~\eqref{eqn:u} and Eq.~\eqref{eqn:v} and taking into account Eqs.~\eqref{eqn:curvature}, Eq.~\eqref{eq5GA} and Eq.~\eqref{eq7GAa} for the bending response, as well as Eq.~\eqref{eqn:eta} and Eq.~\eqref{eq11GA} for the axial response, the whole set of response variables can be cast as functions of the integration constants $\vect{c}_a=[c_{a,1}\;c_{a,2}]^T$ and $\vect{c}_b=[c_{b,1}\;c_{b,2}\;c_{b,3}\;c_{b,4}]^T$, i.e.
\begin{equation}\label{eqn:gen_forces}
\begin{aligned}
&U=U(\xi,\vect{c}_a)\\
&N = -\frac{EA\cdot \lambda^2}{L} \left(\mtder{3}{U}{\xi} - \frac{1}{\lambda^2}\tder{U}{\xi} \right) = N(\xi,\vect{c}_a)\\
&V=V(\xi,\vect{c}_b)\\
&\Theta=-\tder{V}{\xi}=\Theta(\xi,\vect{c}_b)\\
&M  = \frac{EI \cdot \lambda^2}{L^2} \left(\mtder{4}{V}{\xi} - \frac{1}{\lambda^2}\mtder{2}{V}{\xi} \right) = M(\xi,\vect{c}_b)\\
&T = \frac{EI\cdot \lambda^2}{L^3} \left(\mtder{5}{V}{\xi} - \frac{1}{\lambda^2}\mtder{3}{V}{\xi}\right) = T(\xi,\vect{c}_b)\\
\end{aligned}\refstepcounter{equation}\tag{\theequation a-f}
\end{equation}
Computing Eqs.~\eqref{eqn:gen_forces} at $\xi = 0$ and $\xi=1$, the following expressions are obtained for the vectors of nodal displacements and forces:
\begin{align}
&\vect{U}=\vect{A}\vect{c}
\label{eq3GF}\\
&\vect{F}=\vect{B}\vect{c}
\label{eq4GF}
\end{align}
where $\vect{c} = [\vect{c}_a\;\vect{c}_b]^{\mathrm{T}}$. Next, using Eq. \eqref{eq3GF} to calculate $\vect{c}=\vect{A}^{-1} \vect{U}$ and replacing for $\vect{c}$ in Eq. \eqref{eq4GF} lead to \citep{banerjee1997dynamic}
\begin{equation}
\vect{B}\vect{A}^{-1} \vect{U}=\vect{D} \vect{U} = \vect{F}
\label{eq5GF}
\end{equation}
where
\begin{equation}\label{eqn:dsm}
\vect{D}(\omega) = \begin{bmatrix}
\vect{D}_a(\overline{\omega}_a(\omega))&\vect{0}\\
\vect{0}&\vect{D}_b(\overline{\omega}_b(\omega))\\
\end{bmatrix}
\end{equation}
being $\vect{D}_a$ and $\vect{D}_b$ the block matrices associated with axial and bending responses, respectively. The matrix $\vect{D}$ in Eq.~\eqref{eqn:dsm} is the dynamic stiffness matrix of the two-node stress-driven nonlocal beam element in Fig.~\ref{Fig2}. Remarkably, it is available in a closed analytical form, as the inverse matrix $\vect{A}^{-1}$ in Eq.~\eqref{eq5GF} can be obtained symbolically from the inverses of the two separate block matrices associated with axial and bending responses \citep{FAILLA2016171}. The matrix $\vect{D}$ is exact, because is based on the exact solutions of the equations of motion \eqref{eq16GA} and \eqref{eq17GA} along with the related constitutive BCs \eqref{eqn:dimensionless_gen_strain} and \eqref{eqn:dimensionless_curvature}. Indeed, no approximations have been made in building the solutions \eqref{eqn:u} and \eqref{eqn:v}. \\

An alternative approach to derive the exact dynamic stiffness matrix of the two-node stress-driven nonlocal beam element in Fig.~\ref{Fig3} relies on the principle of virtual work.
Consider Eq.~\eqref{eq16GA} governing the steady-state axial vibrations under harmonic axial forces at the beam ends (see Fig.~\ref{Fig3}). The identity of internal and external works reads ($\overline{\omega}_a$\,--\,dependence of response variables and virtual axial displacement/longitudinal generalized strain is omitted for brevity)
\begin{equation}\label{eqn:plv3}
 L \int_{0}^{1}\left(-\lambda^2 \mtder{3}{U}{\xi} + \tder{U}{\xi}\right)\delta\Gamma ~\D{\xi} - \overline{\omega}_{a}^{2} L \int_{0}^{1} U \delta U ~\D{\xi}  = \widetilde{F}_{a,1}\delta U_1 + \widetilde{F}_{a,2}\delta U_2
\end{equation}
where $\delta U_k$ for $k=1,2$ are virtual nodal displacements, $\delta U$ and $\delta \Gamma = \ltder{\delta U}{\xi}$ the corresponding virtual axial displacement and longitudinal generalized strain along the beam; further, $\widetilde{F}_{a,k} = (L^2/EA)F_{a,k}$ for $k=1,2$. Now, exact expression of the axial displacement $U$ is
\begin{equation}\label{eqn:plv_axial_linear}
U(\xi) = \sum_{j=1}^{2}U_{j}\psi_{j}(\xi)
\end{equation}
where $U_j$ are the nodal displacements associated with the applied nodal forces $\widetilde{F}_{a,k}$ and $\psi_j$ are fre\-quen\-cy-de\-pen\-dent, exact shape functions obtained from Eq.~\eqref{eqn:u} enforcing the following BCs (omitting $\overline{\omega}_a$\,--\,dependence for brevity)
\begin{equation}\label{eqn:BC_axial}
\begin{aligned}
&U(0) = 1 \qquad U(1) = 0& &\rightarrow&&\psi_{1}(\xi)\\
&U(0) = 0 \qquad U(1) = 1& &\rightarrow&&\psi_{2}(\xi)\\
\end{aligned}
\end{equation}
The boundary value problem \eqref{eqn:BC_axial} requires inverting a $2\times 2$ matrix to calculate the integration constants $c_{a,k}$ in Eq.~\eqref{eqn:u} and the matrix inversion can be readily implemented in a closed form. Replacing Eq.~\eqref{eqn:plv_axial_linear} in Eq.~\eqref{eqn:plv3} and enforcing the identity of internal and external works for any virtual nodal displacements leads to
\begin{equation}\label{eqn:plv5}
\begin{aligned}
&\sum_{j=1}^{2}\left( L \int_{0}^{1}\left(-\lambda^2 \tder{\psi_i}{\xi}\mtder{3}{\psi_j}{\xi} + \tder{\psi_i}{\xi}\tder{\psi_j}{\xi}\right) ~\D{\xi}\right)U_j - \sum_{j=1}^{2}\left(\overline{\omega}_{a}^{2} L \int_{0}^{1} \psi_i \psi_j ~\D{\xi}\right)U_j = \widetilde{F}_{a,1}\psi_i(0)\\ 
&+ \widetilde{F}_{a,2}\psi_i(1)
\end{aligned}
\end{equation}
where $\psi_{1}(1) = \psi_{2}(0) = 0$ on the r.h.s. Eq.~\eqref{eqn:plv5} can be written in matrix form as
\begin{equation}
\widetilde{\vect{D}}_{a}(\overline{\omega}_a)\vect{U}_{a} = \widetilde{\vect{F}}_{a}
\end{equation}
where $ \vect{U}_{a}=
\begin{bmatrix}
U_{1}&U_{2}
\end{bmatrix}^{\mathrm{T}}$, $\widetilde{\vect{F}}_{a}= \begin{bmatrix}
\widetilde{F}_{a,1}&\widetilde{F}_{a,2}\end{bmatrix}^{\mathrm{T}}$ and $\widetilde{\vect{D}}_a$ is a $2\times 2$ matrix with elements
\begin{equation}\label{eqn:dynamic_stiffness_matrix_rod}
(\widetilde{\vect{D}}_{a})_{ij} = L \int_{0}^{1}\left(-\lambda^2 \tder{\psi_i}{\xi}\mtder{3}{\psi_j}{\xi} + \tder{\psi_i}{\xi}\tder{\psi_j}{\xi}\right)\D{\xi}  - \overline{\omega}_{a}^{2} L \int_{0}^{1} \psi_i \psi_j ~\D{\xi}
\end{equation}
Being $\widetilde{\vect{F}}_{a} = (L^2/EA)\vect{F}_{a}$ for $\vect{F}_{a}= \begin{bmatrix}
F_{a,1}&F_{a,2}\end{bmatrix}^{\mathrm{T}}$, we finally obtain
\begin{equation}
\vect{D}_{a}(\overline{\omega}_a)\vect{U}_{a} = \vect{F}_{a}
\end{equation}
where $\vect{D}_{a}$ is the dynamic stiffness matrix of the rod.
Next, consider Eq.~\eqref{eq17GA} governing the steady-state bending vibrations under transverse forces/moments applied at the beam ends (Fig.~\ref{Fig3}). The identity of internal and external works reads ($\overline{\omega}_b$\,--\,dependence of response variables and virtual deflection/curvature is omitted for conciseness)
\begin{equation}\label{eqn:plvflx3}
\begin{aligned}
&L\int_{0}^{1} \left(\lambda^2 \mtder{4}{V}{\xi} -\mtder{2}{V}{\xi}\right)\delta X \D{\xi}  - \overline{\omega}_{b}^{4} L \int_{0}^{1} V \delta V ~\D{\xi} =\widetilde{F}_{b,1}\delta V_1 + \widetilde{F}_{b,2}\delta V_2 + \widetilde{F}_{b,3}\delta V_3 +\widetilde{F}_{b,4}\delta V_4
\end{aligned}
\end{equation}
where $\delta V_k$ for $k=1,...,4$ are virtual nodal displacements, $\delta V$ and $\delta X=-\lmtder{2}{\delta V}{\xi}$ the corresponding virtual deflection and curvature along the beam; further, $\widetilde{F}_{b,k} = (L^4/EI)F_{b,k}$ for $k=1,3$ and $\widetilde{F}_{b,k} = (L^3/EI)F_{b,k}$ for $k=2,4$, hence:
\begin{equation}\label{eqn:transformation}
\renewcommand{\arraystretch}{1.3}
\underbrace{\begin{bmatrix}
\widetilde{F}_{b,1}\\
\widetilde{F}_{b,2}\\
\widetilde{F}_{b,3}\\
\widetilde{F}_{b,4}
\end{bmatrix}}_{\widetilde{\vect{F}}_{b}} = \underbrace{\begin{bmatrix}
\frac{L^{4}}{EI}&0&0&0\\
0&\frac{L^{3}}{EI}&0&0\\
0&0&\frac{L^{4}}{EI}&0\\
0&0&0&\frac{L^{3}}{EI}
\end{bmatrix}}_{\vect{T}}
\underbrace{
\begin{bmatrix}
F_{b,1}\\
F_{b,2}\\
F_{b,3}\\
F_{b,4}\\
\end{bmatrix}}_{\vect{F}_{b}}
\end{equation}
The exact expression of the deflection $V$ is
\begin{equation}\label{eqn:plv_flexural_linear}
\begin{aligned}
&V(\xi) = \sum_{j=1}^{4}V_{j}\beta_{j}(\xi)
\end{aligned}
\end{equation}
$V_j$ are the nodal displacements associated with the applied nodal forces/moments $F_{b,k}$ and $\beta_j$ are frequency-dependent, exact shape functions obtained from Eq.~\eqref{eqn:v} by enforcing the following BCs (again, omitting $\overline{\omega}_b$\,--\,dependence for brevity)
\begin{equation}\label{eqn:BC_flex}
\begin{aligned}
& U(0) = 1 \qquad \Theta(0) = 0 \qquad U(1) = 0 \qquad \Theta(1) = 0    & &\rightarrow&&\beta_{1}(\xi)\\
& U(0) = 0 \qquad \Theta(0) = 1 \qquad U(1) = 0 \qquad \Theta(1) = 0    & &\rightarrow&&\beta_{2}(\xi)\\
& U(0) = 0 \qquad \Theta(0) = 0 \qquad U(1) = 1 \qquad \Theta(1) = 0    & &\rightarrow&&\beta_{3}(\xi)\\
& U(0) = 0 \qquad \Theta(0) = 0 \qquad U(1) = 0 \qquad \Theta(1) = 1    & &\rightarrow&&\beta_{4}(\xi)\\
\end{aligned}
\end{equation}
The boundary value problem \eqref{eqn:BC_flex} requires inverting a $4\times 4$ matrix to calculate the integration constants $c_{b,k}$ in Eq.~\eqref{eqn:v}, and the matrix inversion can be readily implemented in closed form \citep{FAILLA2016171}. Replacing Eq.~\eqref{eqn:plv_flexural_linear} in Eq.~\eqref{eqn:plvflx3} and enforcing the identity of internal and external works for any virtual nodal displacements yields
\begin{equation}\label{eqn:plvflx5}
\begin{aligned}
&\sum_{j=1}^{4}\left(L\int_{0}^{1}-\lambda^2 \mtder{2}{\beta_i }{\xi}\mtder{4}{\beta_{j}}{\xi} +\mtder{2}{\beta_i}{\xi}\mtder{2}{\beta_{j}}{\xi}\D{\xi}\right)V_j  -\sum_{j=1}^{4}  \left(\overline{\omega}_{b}^{4}L \int_{0}^{1} \beta_i \beta_j  ~\D{\xi}\right)V_j =\widetilde{F}_{b,1}\beta_i (0) \\ 
&- \widetilde{F}_{b,2} \left.\tder{\beta_{i}}{\xi}\right\rvert_{\xi=0} + \widetilde{F}_{b,3}\beta_i(1)-\widetilde{F}_{b,4}\left.\tder{\beta_{i}}{\xi}\right\rvert_{\xi=1}
\end{aligned}
\end{equation} 
being $\beta_i(0)= \left.\tder{\beta_{i}}{\xi}\right\rvert_{\xi=0}=0$ and $\beta_i(1)=\left.\tder{\beta_{i}}{\xi}\right\rvert_{\xi=1} =0$ except for $\beta_1 (0) = 1$, $ \left.\tder{\beta_{2}}{\xi}\right\rvert_{\xi=0} = -1$, $\beta_3(1)=1$, $ \left.\tder{\beta_{4}}{\xi}\right\rvert_{\xi=1}=-1$. Eq.~\eqref{eqn:plvflx5} can be written in matrix form as
\begin{equation}
\widetilde{\vect{D}}_{b}(\overline{\omega}_b)\vect{U}_{b} = \widetilde{\vect{F}}_{b}
\end{equation}
where $\vect{U}_{b}=
\begin{bmatrix}
V_{1}&V_{2}&V_{3}&V_{4}
\end{bmatrix}^{\mathrm{T}}$, $\widetilde{\vect{F}}_{b}=
\begin{bmatrix}\widetilde{F}_{b,1}&\widetilde{F}_{b,2}&\widetilde{F}_{b,3}&\widetilde{F}_{b,4}
\end{bmatrix}^{\mathrm{T}}$ and $\widetilde{\vect{D}}_{b}$ is a $4 \times 4$ matrix whose elements are
\begin{equation}\label{eqn:dynamic_stiffness_matrix_beam_adim}
(\widetilde{\vect{D}}_{b})_{ij} = L\int_{0}^{1}-\lambda^2 \mtder{2}{\beta_i }{\xi}\mtder{4}{\beta_{j}}{\xi} +\mtder{2}{\beta_i}{\xi}\mtder{2}{\beta_{j}}{\xi}\D{\xi} - \overline{\omega}_{b}^{4}L \int_{0}^{1} \beta_i \beta_j  ~\D{\xi}
\end{equation}
From Eq.~\eqref{eqn:dynamic_stiffness_matrix_beam_adim} and taking into account Eq.~\eqref{eqn:transformation}, we finally obtain
\begin{equation}\label{eqn:dynamic_stiffness_matrix}
\vect{D}_{b}(\overline{\omega}_b)\vect{U}_{b} = \vect{F}_{b}
\end{equation}
being
\begin{equation}
\vect{D}_{b}(\overline{\omega}_b) = \vect{T}^{-1}\widetilde{\vect{D}}_{b}
\end{equation}
where $\vect{T}^{-1}$ is trivially computable. Now, assembling Eq.~\eqref{eqn:dynamic_stiffness_matrix_rod} and Eq.~\eqref{eqn:dynamic_stiffness_matrix_beam_adim} for axial and bending vibrations yields

\begin{equation}\label{eqn:full_DSM}
\vect{D}\vect{U} = \vect{F} \qquad 
\vect{D}(\omega) =
\begin{bmatrix}
\vect{D}_{a}(\overline{\omega}_a)&\vect{0}\\
\vect{0}&\vect{D}_{b}(\overline{\omega}_b)
\end{bmatrix}
\end{equation}
where
\begin{equation}
\vect{U}=
\begin{bmatrix}
U_1&U_2&V_{1}&V_{2}&V_{3}&V_{4}
\end{bmatrix}^{\mathrm{T}} \qquad \vect{F}=
\begin{bmatrix}  F_{a,1} &F_{a,2} &F_{b,1}&F_{b,2}&F_{b,3}&F_{b,4}
\end{bmatrix}^{\mathrm{T}}\refstepcounter{equation}\tag{\theequation a,b}
\end{equation}
Remarkably, upon calculating the integrals \eqref{eqn:dynamic_stiffness_matrix_rod} and \eqref{eqn:dynamic_stiffness_matrix_beam_adim}  by standard numerical methods, the matrix $\vect{D}$ in Eq.~\eqref{eqn:full_DSM} is found to coincide with the dynamic stiffness matrix $\vect{D}$ in Eq.~\eqref{eqn:dsm}.\\

\noindent
It is noteworthy that all elements of the dynamic stiffness matrix $\vect{D}$ are real. Further, $\vect{D}$ is symmetric. Indeed, performing integration by parts of Eq.~\eqref{eqn:dynamic_stiffness_matrix_rod} yields
\begin{equation}\label{eqn:sym_dsm_axial}
(\widetilde{\vect{D}}_{a})_{ij} =  L \int_{0}^{1}\left(\lambda^2 \mtder{2}{\psi_i}{\xi}\mtder{2}{\psi_j}{\xi} + \tder{\psi_i}{\xi}\tder{\psi_j}{\xi}\right)\D{\xi}  - \overline{\omega}_{a}^{2} L \int_{0}^{1} \psi_i \psi_j ~\D{\xi}  -\left. \lambda^2 L \tder{\psi_i}{\xi}\mtder{2}{\psi_j}{\xi}\right\rvert_{0}^{1}
\end{equation}
The shape functions $\psi_i$ fulfil Eqs.~\eqref{eqn:dimensionless_gen_strain}, i.e.
\begin{equation}\label{eqn:internal_psi}
\left.\mtder{2}{\psi_i}{\xi}\right\rvert_{\xi=0} = \frac{1}{\lambda}\left.\tder{\psi_i}{\xi}\right\rvert_{\xi=0} \qquad \left.\mtder{2}{\psi_i }{\xi}\right\rvert_{\xi=1} = -\frac{1}{\lambda}\left.\tder{\psi_i}{\xi}\right\rvert_{\xi=1}\refstepcounter{equation}\tag{\theequation a,b} 
\end{equation}
Therefore, in view of Eq.~\eqref{eqn:internal_psi} in Eq.~\eqref{eqn:sym_dsm_axial} takes the form
\begin{equation}\label{eqn:sym_dms_axial_final}
\begin{aligned}
(\widetilde{\vect{D}}_{a})_{ij} &= L \int_{0}^{1}\left(\lambda^2 \mtder{2}{\psi_i}{\xi}\mtder{2}{\psi_j}{\xi} + \tder{\psi_i}{\xi}\tder{\psi_j}{\xi}\right)\D{\xi}  - \overline{\omega}_{a}^{2} L \int_{0}^{1} \psi_i \psi_j ~\D{\xi} \\
&+ \lambda L\left(\left.\tder{\psi_i}{\xi}\tder{\psi_j}{\xi}\right\rvert_{\xi= 0}\right.
\left. + \left.\tder{\psi_i}{\xi}\tder{\psi_j}{\xi}\right\rvert_{\xi=1} \right)
\end{aligned}
\end{equation}
Eq.~\eqref{eqn:sym_dms_axial_final} implies that $(\vect{D}_{a})_{ij} = (\vect{D}_{a})_{ji}$, i.e the symmetry of the dynamic stiffness matrix associated with the axial response.

Likewise, performing integration by parts of Eq.~\eqref{eqn:dynamic_stiffness_matrix_beam_adim} yields
\begin{equation}\label{eqn:sym_dsm_flexural}
(\widetilde{\vect{D}}_{b})_{ij} = L\int_{0}^{1}\left(\lambda^2 \mtder{3}{\beta_i }{\xi}\mtder{3}{\beta_{j}}{\xi} +\mtder{2}{\beta_i}{\xi}\mtder{2}{\beta_{j}}{\xi}\right)\D{\xi} - \overline{\omega}_{b}^{4}L \int_{0}^{1} \beta_i \beta_j  ~\D{\xi}-  \left.  \lambda^2 L\mtder{2}{\beta_i}{\xi}\mtder{3}{\beta_i}{\xi}\right\rvert_{0}^{1}
\end{equation}
Again, since the shape functions $\beta_i$ satisfy Eqs.~\eqref{eqn:dimensionless_curvature}, i.e.
\begin{equation}\label{eqn:internal_beta}
\left.\mtder{3}{\beta_i }{\xi}\right\rvert_{\xi=0} = \frac{1}{\lambda}\left.\mtder{2}{\beta_i}{\xi}\right\rvert_{\xi=0} \qquad \left.\mtder{3}{\beta_i }{\xi}\right\rvert_{\xi=1} = -\frac{1}{\lambda}\left.\mtder{2}{\beta_i}{\xi}\right\rvert_{\xi=1} \refstepcounter{equation}\tag{\theequation a,b}
\end{equation}
Eq.~\eqref{eqn:sym_dsm_flexural} can be written as
\begin{equation}\label{eqn:sym_dsm_flexural_final}
\begin{aligned}
(\widetilde{\vect{D}}_{b})_{ij} &= L\int_{0}^{1}\left(\lambda^2 \mtder{3}{\beta_i }{\xi}\mtder{3}{\beta_{j}}{\xi} +\mtder{2}{\beta_i}{\xi}\mtder{2}{\beta_{j}}{\xi}\right)\D{\xi} - \overline{\omega}_{b}^{4}L \int_{0}^{1} \beta_i \beta_j  ~\D{\xi}\\
&+ \lambda L\left(\left.\mtder{2}{\beta_i}{\xi}\mtder{2}{\beta_j}{\xi}\right\rvert_{\xi= 0}\right.
\left. + \left.\mtder{2}{\beta_i}{\xi}\mtder{2}{\beta_j}{\xi}\right\rvert_{\xi=1} \right)
\end{aligned}
\end{equation}
Eq.~\eqref{eqn:sym_dsm_flexural_final} demonstrates that $(\vect{D}_{b})_{ij} = (\vect{D}_{b})_{ji}$, that is the dynamic stiffness matrix associated with the bending response is symmetric.

At this stage, a few remarks are in order.
\begin{rmk}
The dynamic stiffness matrix $\vect{D}$ of the two-node stress-driven beam element in Fig.~\ref{Fig3} can be used to build the global dynamic stiffness matrix of an arbitrarily-shaped frame. For this, a standard finite-element assembly procedure can be implemented. It is noteworthy that every frame member is modelled exactly by a single element. The size of the global dynamic stiffness matrix depends only on the total number of degrees of freedom of the frame nodes (``beam-to-column'' nodes), as no meshing is required within every frame member. 

The global dynamic stiffness matrix is exact because is exact the dynamic stiffness matrix of every frame member. 
The exact natural frequencies and related modes can be calculated by the WW algorithm, using the implementation described in Section~\ref{sec:4}.

\end{rmk}

\begin{rmk}
The frequency response of the frame, acted upon by harmonic forces/moments at the nodes \citep{banerjee1997dynamic}, can be calculated upon inverting the global dynamic stiffness matrix. This can be done numerically, for every frequency of interest.  
\end{rmk}

\begin{rmk}
The proposed framework can be generalized to build the global dynamic stiffness matrix of 3D frames. This requires formulating the exact dynamic stiffness matrix of a two-node stress-driven nonlocal beam element, where each node features six degrees of freedom including torsional rotation. It is noticed that the stress-driven nonlocal constitutive law for torsional behaviour and associated constitutive BCs, formulated by \cite{barretta2018stress}, lead to a partial differential equation governing torsional vibrations that mirror Eq.~\eqref{eq14GA} for axial vibrations. Therefore, under the assumption of small displacements, the exact dynamic stiffness matrix of the two-node, twelve-degree-of-freedom stress-driven nonlocal beam element will involve separate block matrices pertinent to axial, bending and torsional responses. Again, the global dynamic stiffness matrix will be obtainable by a standard finite-element assembly procedure.
\end{rmk}

\begin{rmk}
The global dynamic stiffness matrix of an arbitrarily-shaped truss can be built based on the dynamic stiffness matrix of a two-node rod, which can be readily derived from Eq.~\eqref{eq5GF} upon eliminating rows and columns associated with the bending response. For trusses as well, the exact natural frequencies and modes can be computed by the WW algorithm, as described in Section~\ref{sec:4}. Both 2D and 3D dimensional truss structures can be modelled.

\end{rmk}

\begin{rmk}
The proposed framework represents an exact approach to the dynamics of small-size frames/trusses, where size effects are modelled by the stress-driven nonlocal model. Here, the assumption is that the nonlocality introduces a coupling between the responses at different points that belong to the same frame/truss member, to an extent depending on the internal length $\lambda$. Recognize that this assumption is made also in the previous works on small-size frames/trusses \citep{NUMANOGLU2019103164,hozhabrossadati2020free}, where Eringen's differential model (\citeyear{Eringen19834703}) was used to build stiffness and mass matrices of a two-node nonlocal finite element. 
\end{rmk}

\begin{rmk}
The differential operators $\,\mathcal{L}_{a}\,$ and $\,\mathcal{L}_{b}\,$ in Eq.~\eqref{eq16GA} and Eq.~\eqref{eq17GA}, 
governing axial and bending free vibrations of the two-node stress-driven nonlocal beam element in Fig.~\ref{Fig3}, are self-adjoint. 
That is,
\begin{align}
&\int_{0}^{1}\mathcal{L}_{a}[U_m]U_n~\D{\xi} =\int_{0}^{1}\mathcal{L}_{a}[U_n]U_m~\D{\xi}\label{eqn:La}\\
&\int_{0}^{1}\mathcal{L}_{b}[V_m]V_n~\D{\xi} =\int_{0}^{1}\mathcal{L}_{b}[V_n]V_m~\D{\xi}\label{eqn:Lb}
\end{align}
with $U_m$, $U_n$, $V_m$, $V_n$ eigenfunctions fulfilling the constitutive BCs \eqref{eqn:dimensionless_gen_strain}-\eqref{eqn:dimensionless_curvature} and the static/kinematic BCs. 
Eq.~\eqref{eqn:La} can be demonstrated writing Eq.~\eqref{eq16GA} for the $m^{\mathrm{th}}$ eigenfunction $U_{m}(\xi)$, multiplying by the $n^{\mathrm{th}}$ eigenfunction $U_n(\xi)$ and integrating Eq.~\eqref{eq16GA} over $[0,1]$, performing integration by parts (as to derive Eq.~\eqref{eqn:sym_dsm_axial}) and enforcing the constitutive BCs \eqref{eqn:dimensionless_gen_strain} along with the static/kinematic BCs. Eq.~\eqref{eqn:Lb} can be proven likewise, starting from Eq.~\eqref{eq17GA}. 
Furthermore, the self-adjoint differential operators $\mathcal{L}_{a}$ and $\mathcal{L}_{b}$ feature properly-defined Green's functions, given by: 
\begin{equation}\label{eqn:Ga}
G_{a}(\xi,\xi_{0}) = g_{a,1} + g_{a,2}\xi + g_{a,3} \lambda^2 \mathrm{e}^{-\frac{\xi}{\lambda}} +  g_{a,4} \lambda^2 \mathrm{e}^{\frac{\xi}{\lambda}} + \left[\xi_{0}-\xi + \frac{\lambda}{2}\left(\mathrm{e}^{\frac{\xi-\xi_{0}}{\lambda}}-\mathrm{e}^{-\frac{\xi-\xi_{0}}{\lambda}} \right)\right]\mathcal{H}(\xi-\xi_{0})
\end{equation}
\begin{equation}\label{eqn:Gb}
\begin{aligned}
G_{b}(\xi,\xi_{0}) =&~~ g_{b,1} + g_{b,2}\xi + g_{b,3}\xi^2 + g_{b,4}\xi^3 + g_{b,5}\lambda^4 \mathrm{e}^{-\frac{\xi}{\lambda}} +  g_{b,6} \lambda^4 \mathrm{e}^{\frac{\xi}{\lambda}} + \frac{1}{6}\left[\vphantom{\frac{\xi-\xi_{0}}{\lambda}}-(\xi-\xi_{0})^3 + 6\lambda^2(\xi-\xi_{0})\right.\\
&\left.+6\lambda^3 \sinh\left(\frac{\xi-\xi_{0}}{\lambda}\right)\right]\mathcal{H}(\xi-\xi_{0})
\end{aligned}
\end{equation}
where $g_{a,i}$ (for $i=1,...,4$) and $g_{b,i}$ (for $i=1,...,6$) are integration constants to be evaluated depending on static/kinematic BCs and $\mathcal{H}$ is the unit-step function defined by
\begin{equation}
\mathcal{H}(\xi-\xi_0) = \left\{
\begin{aligned}
&1&&\text{if}~\xi>\xi_0\\
&0&&\text{if}~\xi<\xi_0
\end{aligned}
\right.
\end{equation} 
The Green's functions \eqref{eqn:Ga} and \eqref{eqn:Gb} are real functions, obtained as solutions of the equations:
\begin{align}
&\mathcal{L}_{a}[U] + \delta(\xi-\xi_0)=0\\
&\mathcal{L}_{b}[V] + \delta(\xi-\xi_0)=0
\end{align}
Specifically, Eq.(53) and Eq.(54) are built applying direct and inverse Laplace transform to Eq.(55) and Eq.(56) respectively (for a similar approach, see \citep{wang2007vibration}).

Since $\mathcal{L}_{a}$ and $\mathcal{L}_{b}$ are self-adjoint, the (real) Green's functions are symmetric. 
Accordingly, the free-vibration problem of the two-node stress-driven nonlocal beam element features an infinite sequence of real eigenvalues (natural frequencies) with associated eigenfunctions, which form an infinite system of functions satisfying the orthogonality conditions \citep{courant1953methods}:
\begin{align}
&L\int_{0}^{1}\rho AU_{m}U_{n}\D{\xi}=\delta_{mn}\\
&L\int_{0}^{1}\rho AV_{m}V_{n}\D{\xi}=\delta_{mn}
\end{align}
with $\delta_{mn}$ Kronecker delta.

Existence of an infinite sequence of real natural frequencies and associated eigenfunctions follows also for the free-vibration problem of an arbitrarily-shaped frame whose members are two-node stress-driven nonlocal beam elements. Indeed, as motivated below, the free-vibration problem of such a frame is still governed by self-adjoint differential operators with associated properly-defined, real and symmetric Green’s functions. 
\begin{itemize}
\item[-] As for self-adjointness, see the work by N\'{a}prstek and Fischer (\citeyear{naprstek2015static}) proving that, upon enforcing the classical equilibrium equations at the nodes, the free-vibration problem of an arbitrarily-shaped frame is still governed by self-adjoint differential operators, if the free-vibration problem of every frame member is governed by self-adjoint differential operators. 
\item[-] As for the calculation of the Green's functions, notice that they can be obtained exactly by a static finite-element analysis of the frame, acted upon by a concentrated force applied at an arbitrary point within one of its members. For this purpose, exact static stiffness matrices and exact static load vectors of the two-node stress-driven nonlocal beam elements can be assembled by a standard finite-element assembly procedure. The exact static stiffness matrix can be obtained mirroring the approach here devised for deriving the exact dynamic stiffness matrix. The exact static load vector of the beam element loaded by an arbitrarily-placed concentrated force can be constructed based on the Green's functions \eqref{eqn:Ga}-\eqref{eqn:Gb} using a standard method to evaluate the corresponding nodal forces (e.g., see the procedure by Failla (\citeyear{FAILLA2016171}) for dynamic problems, applicable also in a static framework). The so-computed Green's functions are real and symmetric as a result of self-adjointess.
\end{itemize}

Additionally, it is noteworthy that existence and uniqueness of the frequency response can be demonstrated for forced-vibration problems governed by self-adjoint operators and associated real, symmetric Green functions \citep{courant1953methods}. Specifically, the frequency response exists and is unique if the forcing frequency is not a natural frequency of the system, which avoids resonance. Accordingly, existence and uniqueness hold true also for the frequency response of the two-node stress-driven nonlocal beam element or a frame whose members are two-node stress-driven nonlocal beam elements.

The conclusions drawn above are valid for any BCs and prove the well-posedness of the stress-driven nonlocal formulation for elastodynamic problems, including free- and forced-vibration ones. On the other hand, the well-posedness of the stress-driven nonlocal formulation for elastostatic problems was already proved by Romano and Barretta (\citeyear{ROMANO201714}), for any BCs as well.

In the next Section, natural frequencies and modes of arbitrarily-shaped frames whose members are two-node stress-driven nonlocal beam elements will be calculated by the WW algorithm.
\end{rmk}

\section{Wittrick-Williams algorithm for small-size trusses and frames}\label{sec:4}
It is known that the WW algorithm calculates all the natural frequencies of a frame whose exact global dynamic stiffness matrix $\vect{D}_{G}(\omega)$ is available \citep{wittrick1971,banerjee1997dynamic}. The unique and distinctive feature of the WW algorithm is that all the natural frequencies are obtained exactly, without missing anyone and including multiple ones \citep{williams1970,wittrick1973,williams1986,banerjee1992,SU2015107}. For this reason, the WW algorithm is the benchmark to investigate the exact free-vibration response of frame.

The basis of the WW algorithm is the calculation of the number of natural frequencies $J(\omega)$ below a trial frequency $\omega$, based on which upper and lower bounds can be determined on every target natural frequency and made to approach each other by the bisection method. Specifically, $J(\omega)$ is given as \citep{wittrick1971}
\begin{equation}\label{eqn:j}
J(\omega) = J_{0}(\omega) + s[\vect{D}_{G}(\omega)]
\end{equation}
where  $J_{0}(\omega)$ is the number of natural frequencies of the component frame members with fixed ends (``clamped-clamped'' frequencies); further, $s[\vect{D}_{G}(\omega)]$ is the number of negative entries on the leading diagonal of the upper triangular matrix obtained by applying the Gaussian elimination procedure to $\vect{D}_{G}(\omega)$.

Here, the objective is to apply the WW algorithm for small-size frames where every member is modelled as the two-node stress-driven nonlocal beam element with exact dynamic stiffness matrix \eqref{eqn:dsm}. 
For this purpose, $s[\vect{D}_{G}(\omega)]$ in Eq.~\eqref{eqn:j} can be obtained from the exact global dynamic stiffness matrix $\vect{D}_{G}(\omega)$, built upon assembling the dynamic stiffness matrices \eqref{eqn:dsm} of the frame members by a standard finite-element assembly procedure, see Section~\ref{sec:3}. Further, taking into account that axial and bending vibration responses of the single frame member are uncoupled (see Eq.~\eqref{eqn:dsm}), $J_{0}(\omega)$ in Eq.~\eqref{eqn:j} can be written as
\begin{equation}\label{eqn:j_0_sum}
J_{0}(\omega) = \sum_{k=1}^{n} J^{(a)}_{0,k}(\omega) + J^{(b)}_{0,k}(\omega)
\end{equation}
where $n$ is the number of the frame members, $J^{(a)}_{0,k}(\omega)$, $J^{(b)}_{0,k}(\omega)$ denote the numbers of ``clamped-clamped'' frequencies smaller than $\omega$ of the $k^{\mathrm{th}}$ frame member,  for axial and bending vibrations respectively.

Now, for consistency with the whole formulation in Section~\ref{sec:3}, the calculation of $s[\vect{D}_{G}(\omega)]$, $J^{(a)}_{0,k}(\omega)$ and $J^{(b)}_{0,k}(\omega)$ is performed using dimensionless frequencies $\overline{\omega}_a$ and $\overline{\omega}_b$ corresponding to the trial frequency $\omega$. This is straightforward in the calculation of $s[\vect{D}_{G}(\omega)]$, see Eq.~\eqref{eqn:dsm} for the dynamic stiffness matrix $\vect{D}$ of every frame member. On the other hand, the general expressions for $J^{(a)}_{0,k}(\overline{\omega}_a)$ and $J^{(b)}_{0,k}(\overline{\omega}_b)$ are
\begin{equation}\label{eqn:j_0_value}
J^{(a)}_{0,k}(\overline{\omega}_a) = \sum_{r = 1}^{\infty} \min\left\{1; \left\lfloor \frac{\overline{\omega}_a}{\overline{\omega}^{(a)}_{k,r}}\right\rfloor \right\}; \qquad \qquad J^{(b)}_{0,k}(\overline{\omega}_b) = \sum_{r = 1}^{\infty} \min\left\{1; \left\lfloor \frac{\overline{\omega}_b}{\overline{\omega}^{(b)}_{k,r}}\right\rfloor \right\}\refstepcounter{equation}\tag{\theequation a,b}
\end{equation}
where $\overline{\omega}^{(\diamondsuit)}_{k,r}$ (with $\diamondsuit=a,b$) is the $r^{\mathrm{th}}$ dimensionless ``clamped-clamped'' frequency of the $k^{\mathrm{th}}$ frame member with either axial or bending vibrations ($\overline{\omega}^{(\diamondsuit)}_{k,1} \leq \overline{\omega}^{(\diamondsuit)}_{k,2} \leq \dots \leq  \overline{\omega}^{(\diamondsuit)}_{k,r} \leq  \overline{\omega}^{(\diamondsuit)}_{k,r+1} \leq \dots$). For the stress-driven nonlocal beam under study, $\overline{\omega}^{(\diamondsuit)}_{k,r}$ shall be obtained as roots of a characteristic equation given as determinant of matrix $\vect{A}$ in Eq.~\eqref{eq3GF}. Specifically, two separate characteristic equations can be obtained from the determinants of the block matrices $\vect{A}_a$ and $\vect{A}_b$ associated with axial and bending vibrations:
\begin{align}
&\det{\vect{A}_a (\overline{\omega}_{a})} = g_1(0,\overline{\omega}_b)g_2(1,\overline{\omega}_b)-g_1 (1,\overline{\omega}_b) g_2 (0,\overline{\omega}_b) \label{eqn:det_A}\\
&\det{\vect{A}_b (\overline{\omega}_{b})} = \sum_{i,j,h,k = 1}^{4}g_{i}(0,\overline{\omega}_b) \cdot g_{j}(1,\overline{\omega}_b) \cdot \left.\pder{g_h(\xi,\overline{\omega}_b)}{\xi}\right\rvert_{\xi=0}\cdot\left.\pder{g_k(\xi,\overline{\omega}_b)}{\xi}\right\rvert_{\xi=1}\epsilon_{ijhk} \label{eqn:det_B}
\end{align}
being $\epsilon_{ijhk}$ the $4^{\mathrm{th}}$-dimensional Levi-Civita tensor defined as
\begin{equation}\label{eqn:levi_civita}
\epsilon_{ijhk} = \left\{      
\begin{aligned}
&1&&\text{if}~(i,j,k,h) = (1,2,3,4),(2,3,4,1),(3,4,1,2)~\text{or}~(4,1,2,3)      \\
-&1&&\text{if}~(i,j,k,h) = (4,3,2,1),(3,2,1,4),(2,1,4,3)~\text{or}~(1,4,3,2) \\      &0&&\text{if two or more indices are equal}\end{aligned}      
\right.
\end{equation}

It is noticed that the roots of Eq.~\eqref{eqn:det_A} and Eq.~\eqref{eqn:det_B} cannot be obtained in analytical form but are readily obtainable by a numerical root-finding algorithm. Indeed, whenever roots of the characteristic equation are not available in analytical form, the implementation of the WW algorithm involves the numerical calculation of the roots in order to evaluate $J_0(\omega)$, e.g. see \citep{banerjee1985}. Table~\ref{tbl:dimensionless_frequencies} reports the first 20 dimensionless ``clamped-clamped'' frequencies of the two-node stress-driven nonlocal beam element in Figure~\ref{Fig3}, for various values of the internal length $\lambda$, as computed by Mathematica \citep{Mathematica}. Additional frequencies are not included in Table~\ref{tbl:dimensionless_frequencies} for conciseness.
\begin{table}[pos=ht]
\caption{First 20 dimensionless frequencies of clamped-clamped two-node stress-driven nonlocal beam element: (a) axial vibrations, (b) bending vibrations}\label{tbl:dimensionless_frequencies}
\begin{subtable}{0.45\linewidth}
\begin{tabular}{cccc}
\toprule[1pt]
$\lambda $&0&0.01&0.10\\
\midrule[0.5pt]
 & 3.14159 & 3.17488 & 3.63694 \\ 
 & 6.28319 & 6.35908 & 8.07878 \\ 
 & 9.42478 & 9.56186 & 13.86928 \\ 
 & 12.56637 & 12.79240 & 21.29970 \\ 
 & 15.70796 & 16.05973 & 30.51843 \\ 
 & 18.84956 & 19.37271 & 41.60268 \\ 
 & 21.99115 & 22.73997 & 54.59429 \\ 
 & 25.13274 & 26.16990 & 69.51703 \\ 
 & 28.27433 & 29.67060 & 86.38503 \\ 
 & 31.41593 & 33.24986 & 105.20705 \\ 
\cmidrule[1pt]{2-4}
\end{tabular}
\caption{\label{tab:freq_ass}\qquad\qquad\qquad}
\end{subtable}
\begin{subtable}{0.45\linewidth}
\begin{tabular}{cccc}
\toprule[1pt]
$\lambda $&0&0.01&0.10\\
\midrule[0.5pt]
 & 4.71239 & 4.78036 & 5.46176 \\ 
 & 7.85398 & 7.94458 & 9.61519 \\ 
 & 10.99557 & 11.13962 & 14.37222 \\ 
 & 14.13717 & 14.34978 & 19.73479 \\ 
 & 17.27876 & 17.58029 & 25.67782 \\ 
 & 20.42035 & 20.83533 & 32.16655 \\ 
 & 23.56194 & 24.11894 & 39.16664 \\ 
 & 26.70354 & 27.43498 & 46.64714 \\ 
 & 29.84513 & 30.78710 & 54.58095 \\ 
 & 32.98672 & 34.17872 & 62.94445 \\ 
\cmidrule[1pt]{2-4}
\end{tabular}
\caption{\label{tab:freq_beam}\qquad\qquad\qquad}
\end{subtable}
\end{table}

\break

\section{Numerical applications}\label{sec:5}
Here, two applications are presented. For a first insight, natural frequencies and modes of a single small-size beam will be calculated by the proposed approach and validated by comparison with approximate ones obtained by an alternative solution method. Secondly, the proposed approach will be applied to a small-size structure.

\subsection{Example A}
\vspace{10pt}
\noindent
Consider the small-size cantilever beam in Fig.~\ref{fig:cantilever}, with the following parameters: $E=\SI{427}{\GPa}$, $L=\SI{20}{\nm}$, $A=\SI{2}{\unexp{\nm}{2}}$ (rectangular cross section with $b=\SI{1}{\nm}$ and $h=\SI{2}{\nm}$), $\rho=\SI{3200}{\kg \unexp{\m}{-3}}$. The proposed approach is implemented to investigate axial and bending vibrations. Specifically, the exact dynamic stiffness matrix $\vect{D}(\omega)$ of the beam is constructed using a single element and the WW algorithm described in Section \ref{sec:4} is applied to calculate exact natural frequencies and modes. For validation, they are compared with approximate frequency/modes obtained by a linear eigenvalue problem formulated by the Rayleigh-Ritz method \citep{meirovitch1997principles}, using Chebysh\"{e}v polynomials as trial functions. Details on the implementation of the Rayleigh-Ritz method are given in Appendix B.
{
\vspace{15pt}
\begin{figure}[pos=h]
\centering
\includegraphics[scale=0.33]{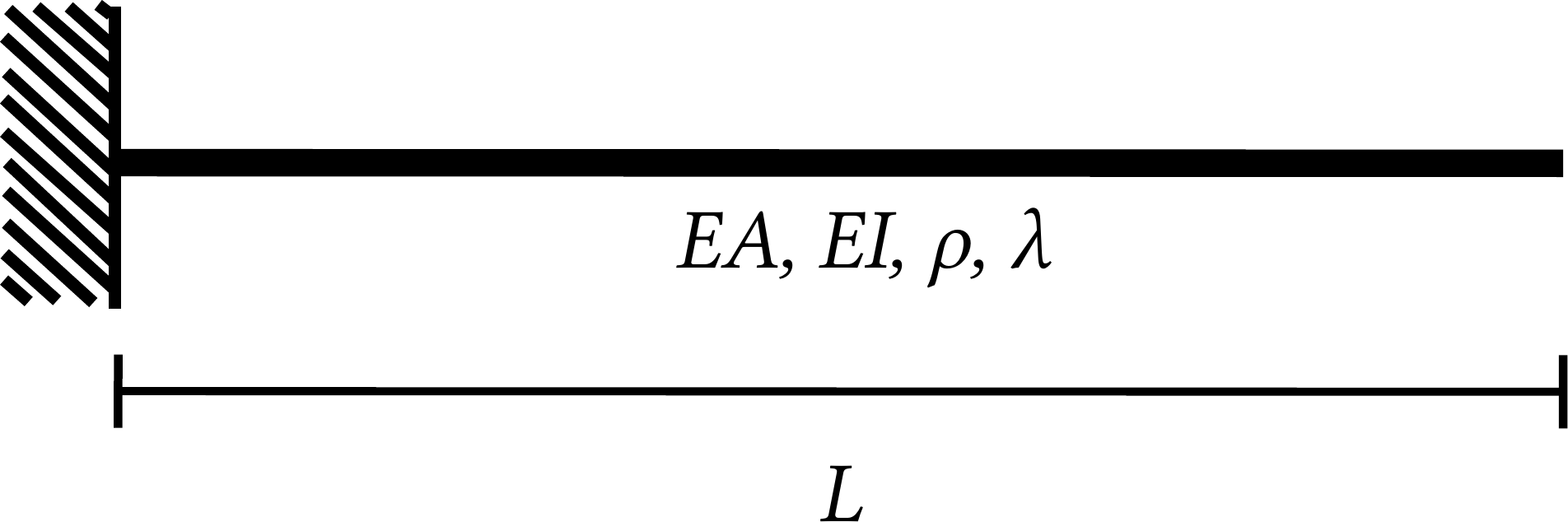}
\caption{\label{fig:cantilever}Small-size cantilever beam.}
\end{figure}
\vspace{15pt}

Tables \ref{tbl:cantilever_axial} and \ref{tbl:cantilever_flexural} report the first $10$ natural frequencies of axial and bending vibrations, assuming $\lambda=0.1$ as internal length of the stress-driven nonlocal model. As the number $N$ of Chebysh\"{e}v polynomials increases, the approximate natural frequencies obtained by the Rayleigh-Ritz method converge to the exact ones calculated by the proposed approach, with accuracy up to the first four digits after the comma. For completeness, Tables \ref{tbl:cantilever_axial} and \ref{tbl:cantilever_flexural} include also the numbers of $J_0$ and $s[\vect{D}]$ corresponding to each natural frequency, as calculated by the WW algorithm.

\begin{table}[pos=h]
\caption{Natural frequencies for axial vibrations of cantilever beam in  Fig.~\ref{fig:cantilever}.}\label{tbl:cantilever_axial}
\begin{tabular}{cccccccc}
\toprule[1pt]
\multicolumn{4}{c}{Rayleigh-Ritz method}&\multicolumn{3}{c}{Proposed approach}\\
\midrule[0.5pt]
$\omega$ (GHz)&$\omega$ (GHz)&$\omega$ (GHz)&$\omega$ (GHz)&\multirow{2}{*}{$\omega$ (GHz)}&\multirow{2}{*}{$J_0$}&\multirow{2}{*}{$s[\vect{D}]$}\\
$N = 15$&$N=20$&$N=25$&$N=30$&&&\\
\midrule[0.5pt]
153.55326&153.55326&153.55326&153.55326&153.55326&0&1\\
496.47072&496.47072&496.47072&496.47072&496.47072&1&1\\
935.15490&935.15490&935.15490&935.15490&935.15490&2&1\\
1507.35332&1507.35332&1507.35332&1507.35332&1507.35332&3&1\\
2234.00705&2234.00701&2234.00701&2234.00701&2234.00701&4&1\\
3126.45565&3126.44056&3126.44056&3126.44056&3126.44056&5&1\\
4191.19849&4190.85881&4190.85880&4190.85880&4190.85880&6&1\\
5446.24457&5430.76615&5430.76561&5430.76561&5430.76561&7&1\\
6923.28272&6848.28865&6848.21510&6848.21510&6848.21510&8&1\\
9171.87686&8445.24453&8444.45914&8444.45902&8444.45902&9&1\\
\bottomrule[1pt]
\end{tabular}
\end{table}

\begin{table}[pos=h]
\caption{Natural frequencies for bending vibrations of cantilever beam in Fig.~\ref{fig:cantilever}.}\label{tbl:cantilever_flexural}
\begin{tabular}{cccccccc}
\toprule[1pt]
\multicolumn{4}{c}{Rayleigh-Ritz method}&\multicolumn{3}{c}{Proposed approach}\\
\midrule[0.5pt]
$\omega$ (GHz)&$\omega$ (GHz)&$\omega$ (GHz)&$\omega$ (GHz)&\multirow{2}{*}{$\omega$ (GHz)}&\multirow{2}{*}{$J_0$}&\multirow{2}{*}{$s[\vect{D}]$}\\
$N = 15$&$N=20$&$N=25$&$N=30$&&&\\
\midrule[0.5pt]
10.34411&10.34411&10.34411&10.34411&10.34411&0&1\\
69.34614&69.34614&69.34614&69.34614&69.34614&0&2\\
216.98244&216.98244&216.98244&216.98244&216.98244&1&2\\
486.95413&486.95413&486.95413&486.95413&486.95413&2&2\\
924.34254&924.34242&924.34242&924.34242&924.34242&3&2\\
1576.81346&1576.71498&1576.71498&1576.71498&1576.71497&4&2\\
2493.86913&2492.72293&2492.72281&2492.72281&2492.72281&5&2\\
3777.69185&3721.45057&3721.44739&3721.44739&3721.44738&6&2\\
5522.11124&5312.63783&5312.14577&5312.14577&5312.14575&7&2\\
9394.99642&7317.91183&7314.14894&7314.14767&7314.14765&8&2\\
\bottomrule[1pt]
\end{tabular}
\end{table}

Figures \ref{fig:axial_eigenfunctions} and \ref{fig:flexural_eigenfunctions} show the eigenfunctions corresponding to the first four modes for axial and bending vibrations, calculated by the proposed approach and Rayleigh-Ritz method. The agreement is excellent, substantiating the correctness of the proposed strategy.

\begin{figure}[pos=ht]
\centering
\includegraphics[scale=0.8]{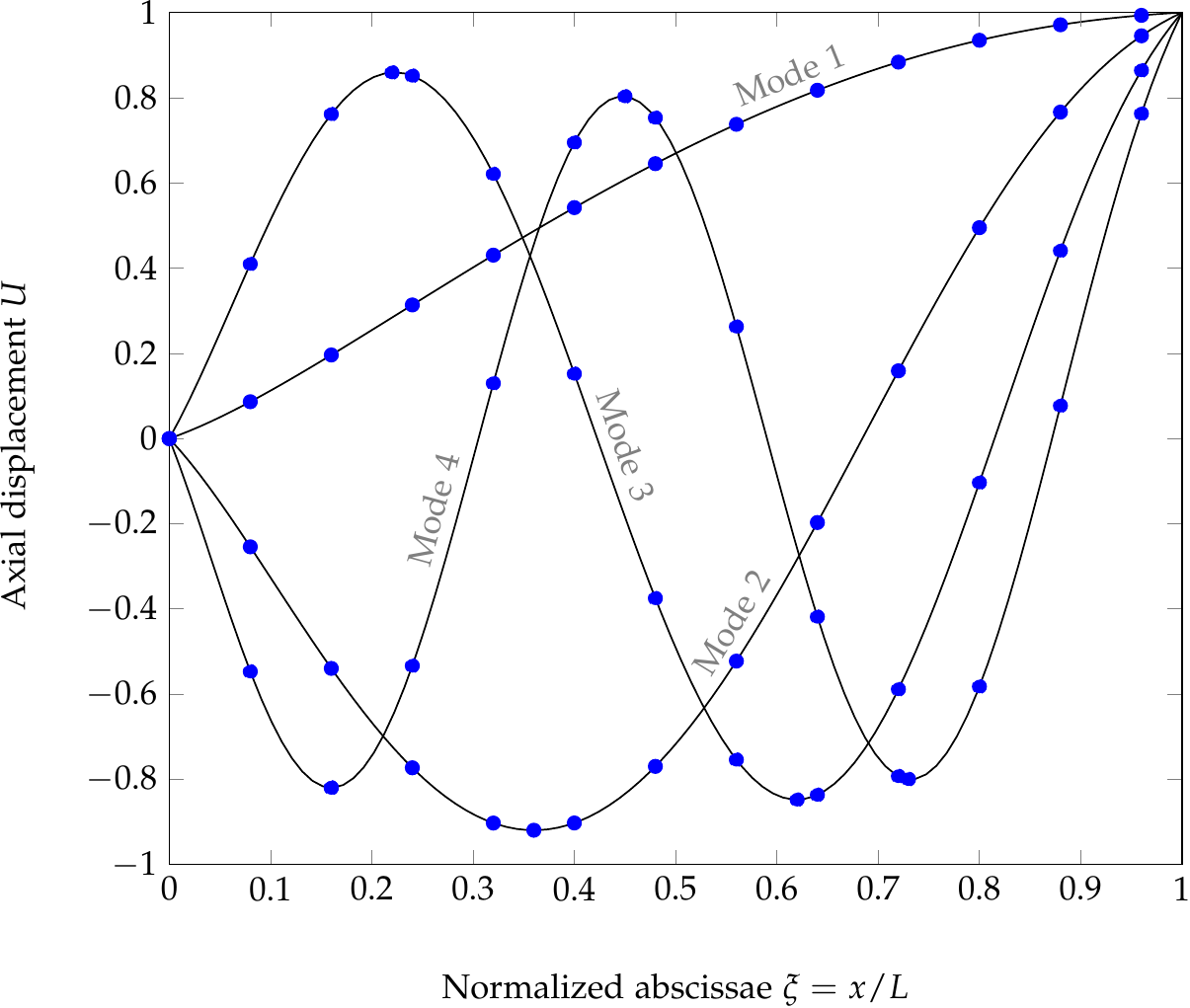}
\caption{\label{fig:axial_eigenfunctions}First four eigenfunctions for axial vibrations of the cantilever beam in Fig.~\ref{fig:cantilever}: Proposed approach (continuous line), Rayleigh-Ritz method (dots).}
\end{figure}
\begin{figure}[pos=ht]
\centering
\includegraphics[scale=0.8]{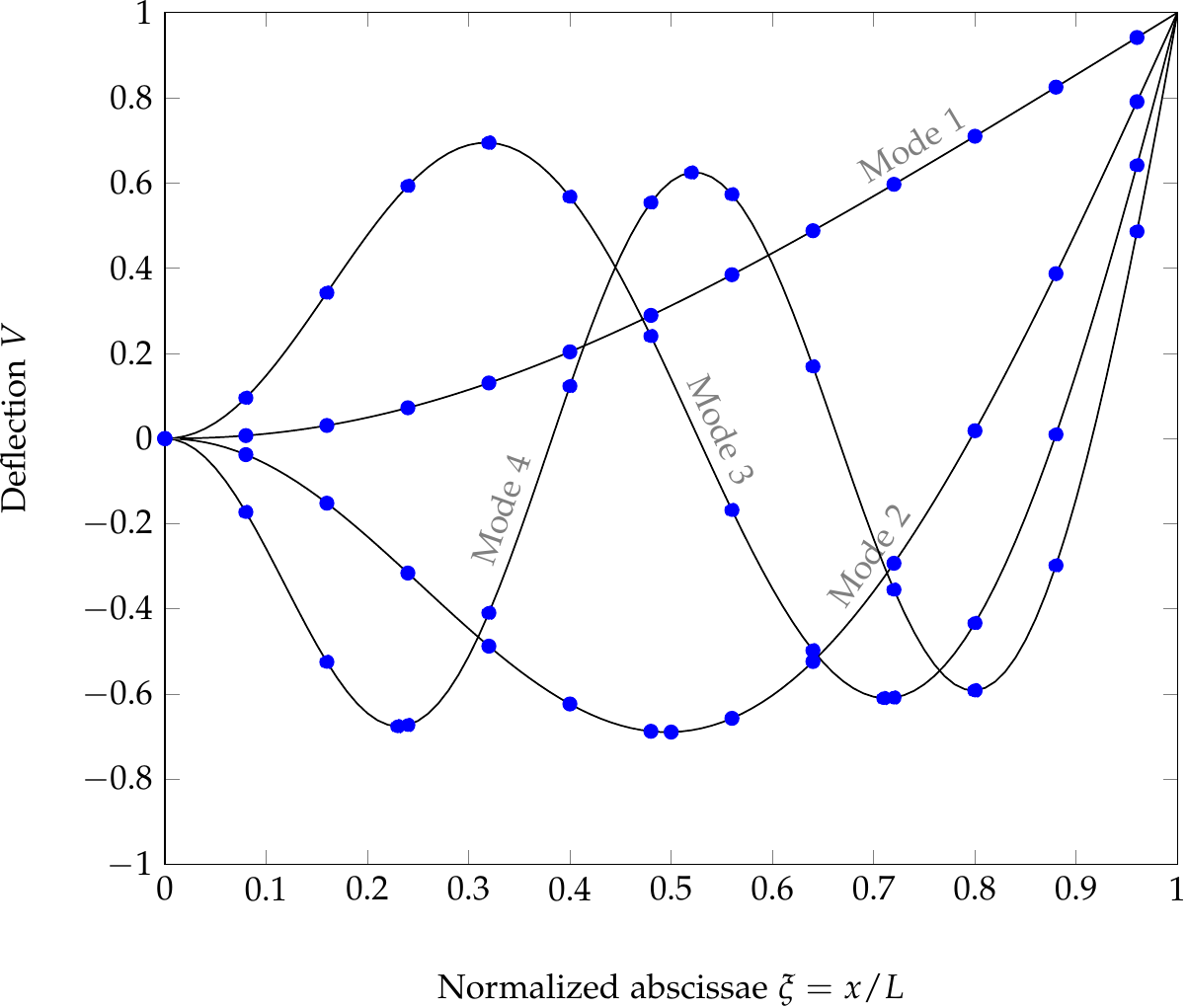}
\caption{\label{fig:flexural_eigenfunctions}First four eigenfunctions for bending vibrations of the cantilever beam in Fig.~\ref{fig:cantilever}: Proposed approach (continuous line), Rayleigh-Ritz method (dots).}
\end{figure}
}
\clearpage
\subsection{Example B}
\begin{figure}[pos=h]
\centering
\includegraphics[scale=0.22]{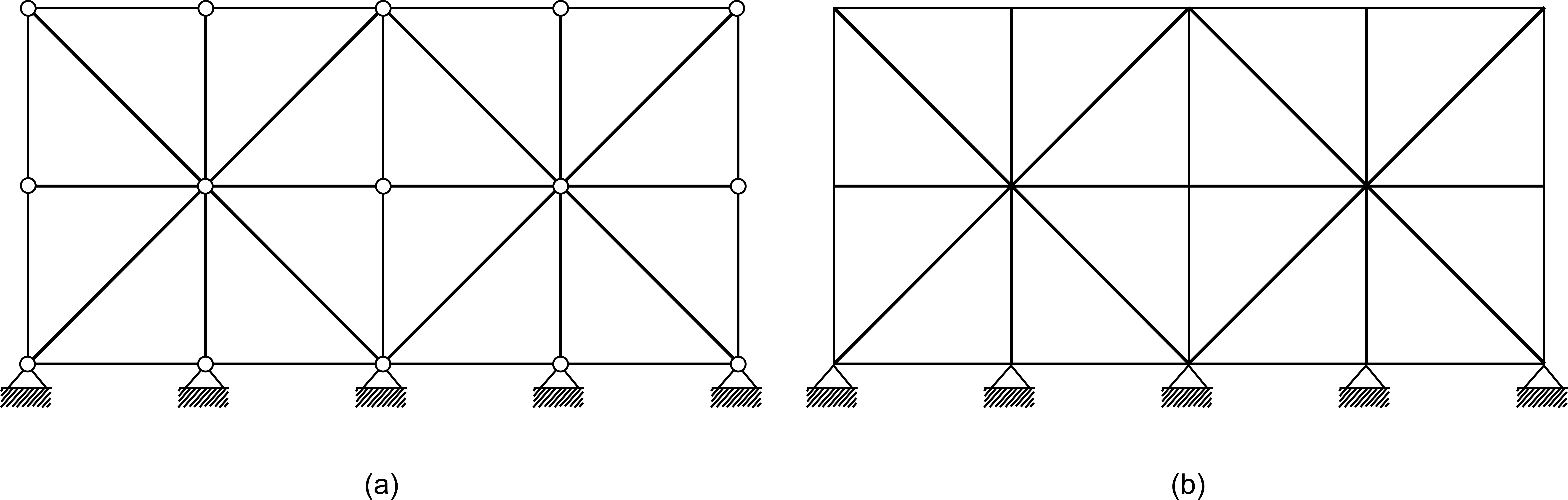}
\caption{\label{fig:frame} Small-size 2D structures: (a) truss; (b) frame.}
\end{figure}

Consider the small-size 2D structures in Fig.~\ref{fig:frame}. The two structures feature the same external constraints but two different assumptions are made on the internal nodes, i.e. truss (Fig.~\ref{fig:frame}a) and frame (Fig.~\ref{fig:frame}b) nodal connections are considered. For numerical purposes, reference parameters are: $E=\SI{427}{\GPa}$, $L=\SI{20}{\nm}$, $A=\SI{2}{\unexp{\nm}{2}}$ (rectangular cross section with $b=\SI{1}{\nm}$ and $h=\SI{2}{\nm}$), $\rho=\SI{3200}{\kg \unexp{\m}{-3}}$. 
To investigate the free-vibration response in presence of size effects, every structural member is modelled by a single stress-driven nonlocal two-node beam element, whose exact dynamic stiffness matrix is $\vect{D}$ given by Eq.~\eqref{eqn:dsm} for the frame or its block matrix $\vect{D}_a$ for the truss. On building the global dynamic stiffness matrix by a standard finite-element assembly procedure, exact natural frequencies and modes are calculated by the WW algorithm described in Section~\ref{sec:4}. Size effects are investigated considering different values of the internal length $\lambda$ of the stress-driven nonlocal model.

\begin{table}[pos=h]
\caption{Natural frequencies, $J_{0}$ and $s[\vect{D}_{G}]$ for truss in Fig.~\ref{fig:frame}a.}\label{tbl:truss}
\begin{tabular}{cccccccccc}
\toprule[1pt]
$\lambda $&\multicolumn{3}{c}{0}&\multicolumn{3}{c}{0.01}&\multicolumn{3}{c}{0.10}\\
\midrule[0.5pt]
&$\omega~(\SI{}{\GHz})$&$J_0$&$s[\vect{D}_{G}]$&$\omega~(\SI{}{\GHz})$&$J_0$&$s[\vect{D}_{G}]$&$\omega~(\SI{}{\GHz})$&$J_0$&$s[\vect{D}_{G}]$\\
\cmidrule[0.5pt](lr{0.125em}){2-4}\cmidrule[0.5pt](lr{0.125em}){5-7}\cmidrule[0.5pt](lr{0.125em}){8-10}
 & 27.43821 & 0 & 1 & 27.57524 & 0 & 1 & 28.92544 & 0 & 1 \\ 
 & 58.27909 & 0 & 2 & 58.56994 & 0 & 2 & 61.51358 & 0 & 2 \\ 
 & 63.87465 & 0 & 4 & 64.20284 & 0 & 4 & 67.53415 & 0 & 4 \\ 
 & 63.87465 & 0 & 4 & 64.20284 & 0 & 4 & 67.53415 & 0 & 4 \\ 
 & 72.19697 & 0 & 5 & 72.56269 & 0 & 5 & 76.27531 & 0 & 5 \\ 
 & 75.49053 & 0 & 6 & 75.86652 & 0 & 6 & 79.78619 & 0 & 6 \\ 
 & 89.90719 & 0 & 7 & 90.34393 & 0 & 7 & 94.99763 & 0 & 7 \\ 
 & 101.00278 & 0 & 9 & 101.47942 & 0 & 9 & 106.72088 & 0 & 9 \\ 
 & 101.00278 & 0 & 9 & 101.47942 & 0 & 9 & 106.72088 & 0 & 9 \\ 
 & 130.70334 & 0 & 10 & 131.39556 & 0 & 10 & 139.26081 & 0 & 10 \\ 
 & 144.39394 & 0 & 11 & 145.13719 & 0 & 11 & 153.55326 & 0 & 11 \\ 
 & 161.97629 & 0 & 13 & 162.98630 & 0 & 13 & 174.80492 & 0 & 13 \\ 
 & 161.97629 & 0 & 13 & 162.98630 & 0 & 13 & 174.80492 & 0 & 13 \\ 
 & 170.78182 & 0 & 14 & 171.94402 & 0 & 14 & 185.74609 & 0 & 14 \\ 
 & 176.32509 & 0 & 15 & 177.47886 & 0 & 15 & 191.44679 & 0 & 15 \\ 
 & 184.61555 & 0 & 17 & 185.96370 & 0 & 17 & 202.32571 & 0 & 17 \\ 
 & 184.61555 & 0 & 17 & 185.96370 & 0 & 17 & 202.32571 & 0 & 17 \\ 
 & 191.85137 & 0 & 18 & 193.44083 & 0 & 18 & 212.98799 & 0 & 18 \\ 
 & 209.44418 & 8 & 11 & 211.20072 & 8 & 11 & 232.47553 & 0 & 19 \\ 
 & 216.59091 & 8 & 12 & 217.73242 & 8 & 12 & 234.52643 & 0 & 20 \\ 
 \cmidrule[1pt]{2-10}
\end{tabular}
\end{table}

\break
For a first insight, the truss model in Fig.~\ref{fig:frame}a is considered. Table \ref{tbl:truss} shows the first 20 natural frequencies computed by the proposed approach, along with the numbers $J_0$ and $s[\vect{D}_{G}]$ corresponding to each natural frequency calculated by the WW algorithm. Results in Table \ref{tbl:truss} suggest some interesting comments. The first is that the natural frequencies increase with the internal length $\lambda$, meaning that increasing nonlocality induces stiffening. This is in agreement with previous results obtained for the static response using the stress-driven nonlocal approach \citep{ROMANO201714,ROMANO2017184,ROMANO2017151}; as for size effects in general, it is to be noticed that experimental evidence of stiffening size effects exists in the literature for small-size specimens, see for instance the work by \cite{LAM20031477}. A second observation is that the natural frequencies tend to those of the classical local model as the internal length $\lambda$ decreases, as expected.

\begin{figure}[pos=h]
\centering
\begin{subfigure}[pos=b]{.46\linewidth}
\raisebox{-37.8mm}{\includegraphics[scale=0.32]{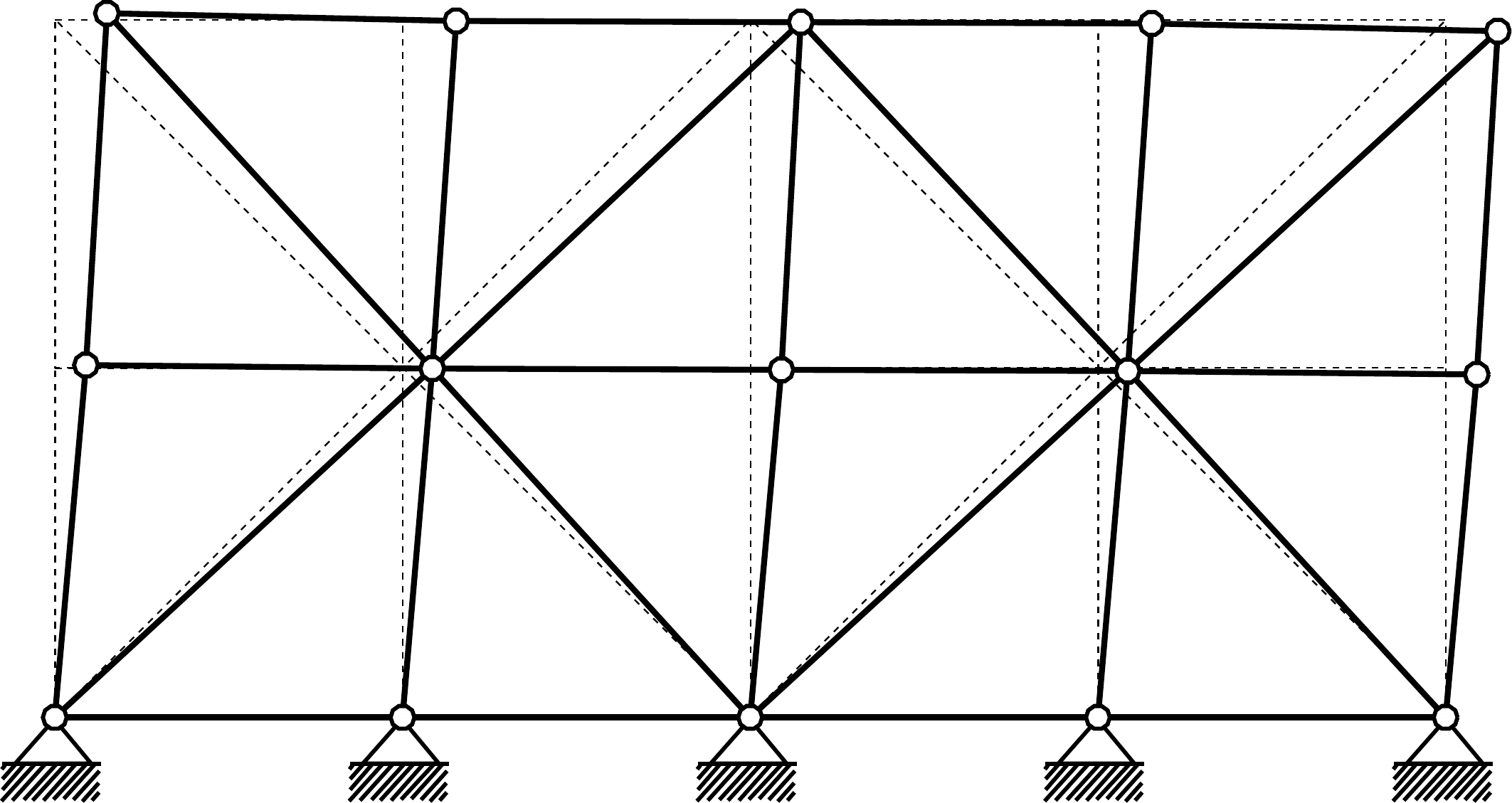}}
\caption{\label{fig:mode_1_ax}$1^{\mathrm{st}}$ mode}
\end{subfigure}
\begin{subfigure}[pos=t]{.46\linewidth}
\includegraphics[scale=0.32]{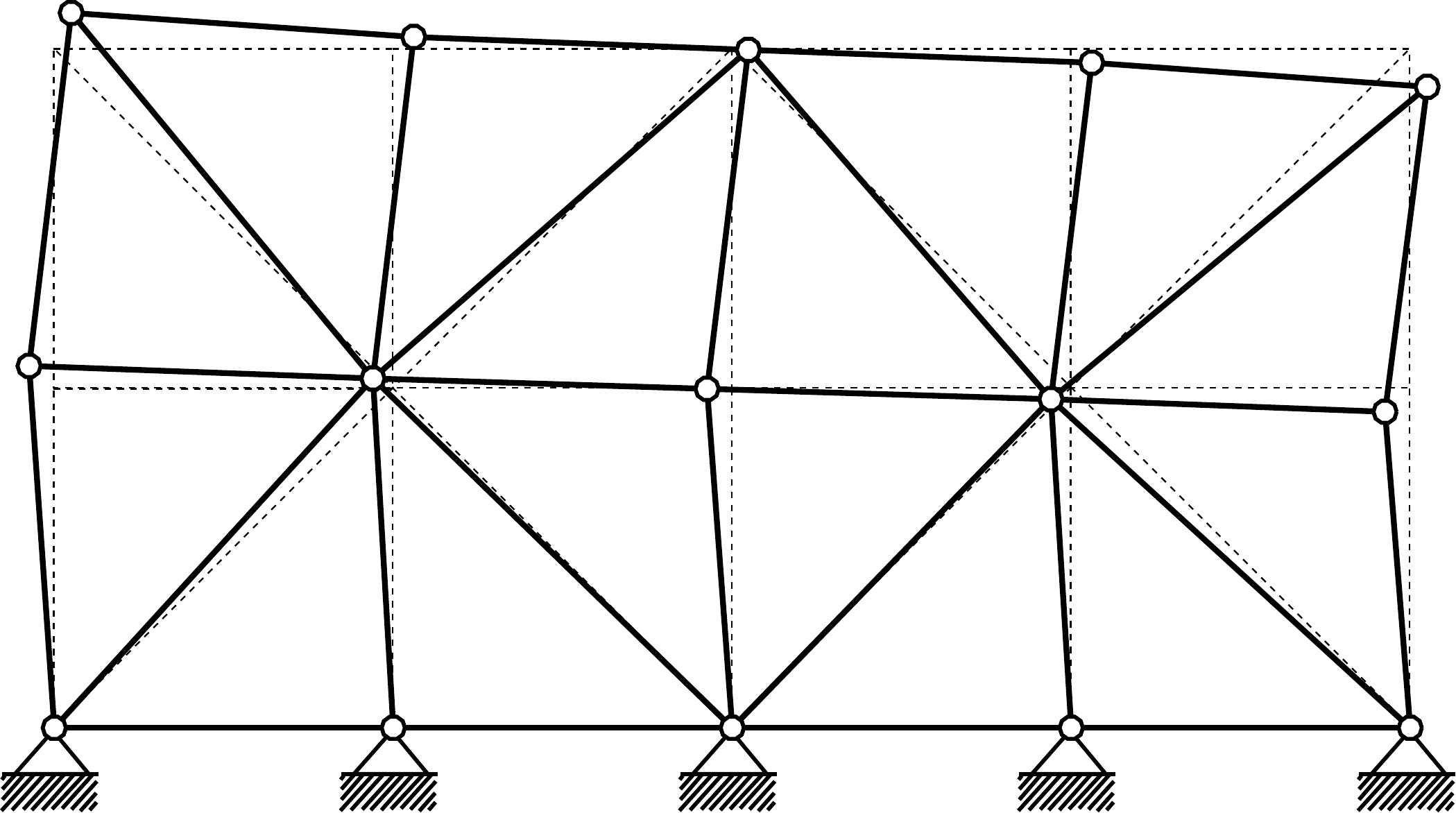}
\caption{\label{fig:mode_2_ax}$2^{\mathrm{nd}}$ mode}
\end{subfigure}\\\vspace{1cm}
\begin{subfigure}[pos=t]{.46\linewidth}
\raisebox{-40.7mm}{\includegraphics[scale=0.32]{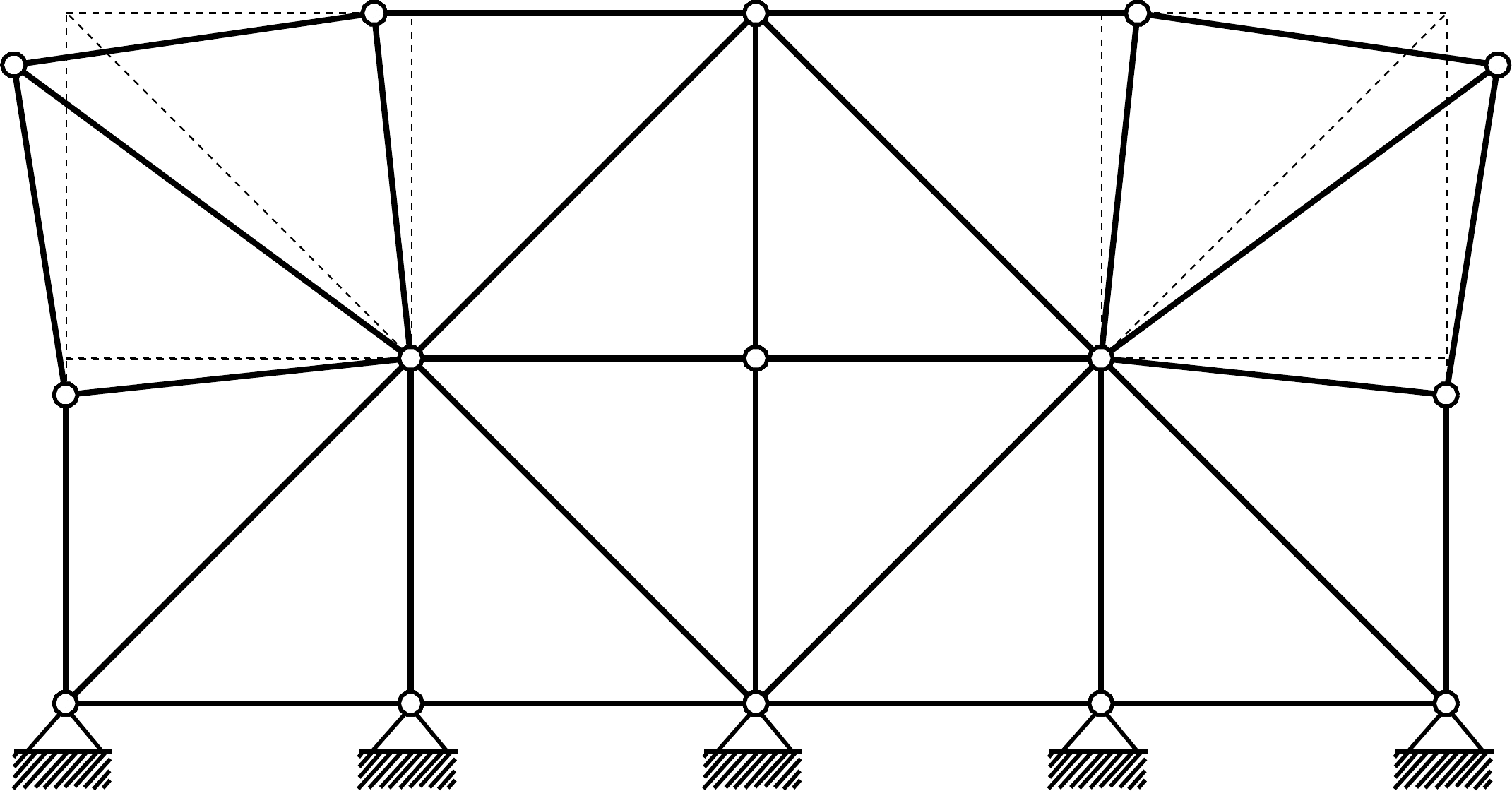}}
\caption{\label{fig:mode_5_ax}$5^{\mathrm{th}}$ mode}
\end{subfigure}
\begin{subfigure}[pos=t]{.46\linewidth}
\includegraphics[scale=0.32]{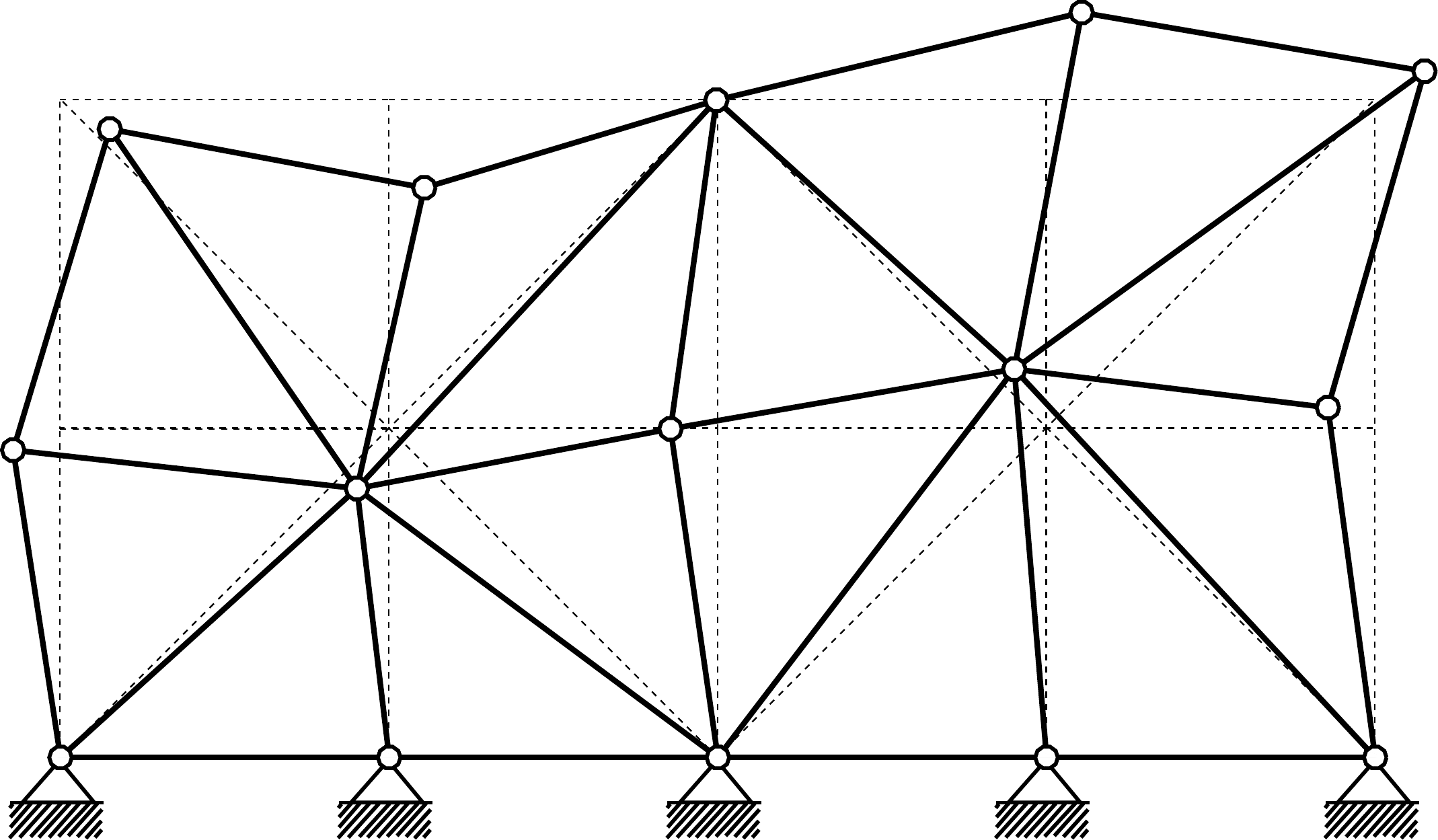}
\caption{\label{fig:mode_6_ax}$6^{\mathrm{th}}$ mode}
\end{subfigure}
\caption{\label{fig:vibration_mode}Mode shapes for truss in Fig.~\ref{fig:frame}a, for $\lambda = 0.10$.}
\end{figure}

Finally, it is noteworthy that some of the natural frequencies reported in Table~\ref{tbl:truss} are double roots of the characteristic equation, confirming that the WW algorithm is capable of detecting all natural frequencies, including multiple ones.
The mode shapes associated with some of the natural frequencies in Table~\ref{tbl:truss} are illustrated in Fig.~\ref{fig:vibration_mode}. 
They appear meaningful based on engineering judgement and exhibit typical symmetric or anti-symmetric shapes, as is typical the case in vibrating structures.

Next, the frame in Fig.~\ref{fig:frame}b is investigated. Table~\ref{tbl:beams} reports the first 20 natural frequencies obtained by the proposed approach along with the pertinent numbers $J_0$ and $s[\vect{D}_{G}]$. Comments mirrors those on Table~\ref{tbl:truss}. That is, the natural frequencies increase with the internal length $\lambda$, i.e. increasing nonlocality induces stiffening effects in the free-vibration response, while the local natural frequencies are retrieved for vanishing $\lambda$. For a final insight, Fig.~\ref{fig:FRF} illustrates the frequency response for the horizontal displacement of the top-right node, when a harmonic horizontal force $1e^{i\omega t}$ (expressed in \SI{}{\nN}) is applied at the same node of the frame. As expected, the resonance peaks of the frequency response occur at the natural frequencies reported in Table~\ref{tbl:beams} for various internal lengths $\lambda$'s. The results in Fig.~\ref{fig:FRF} substantiate the correctness of the proposed approach.

\begin{figure}[pos=h]
\centering
\begin{subfigure}[pos=b]{.45\linewidth}
\raisebox{0.5mm}{\includegraphics[scale=0.32]{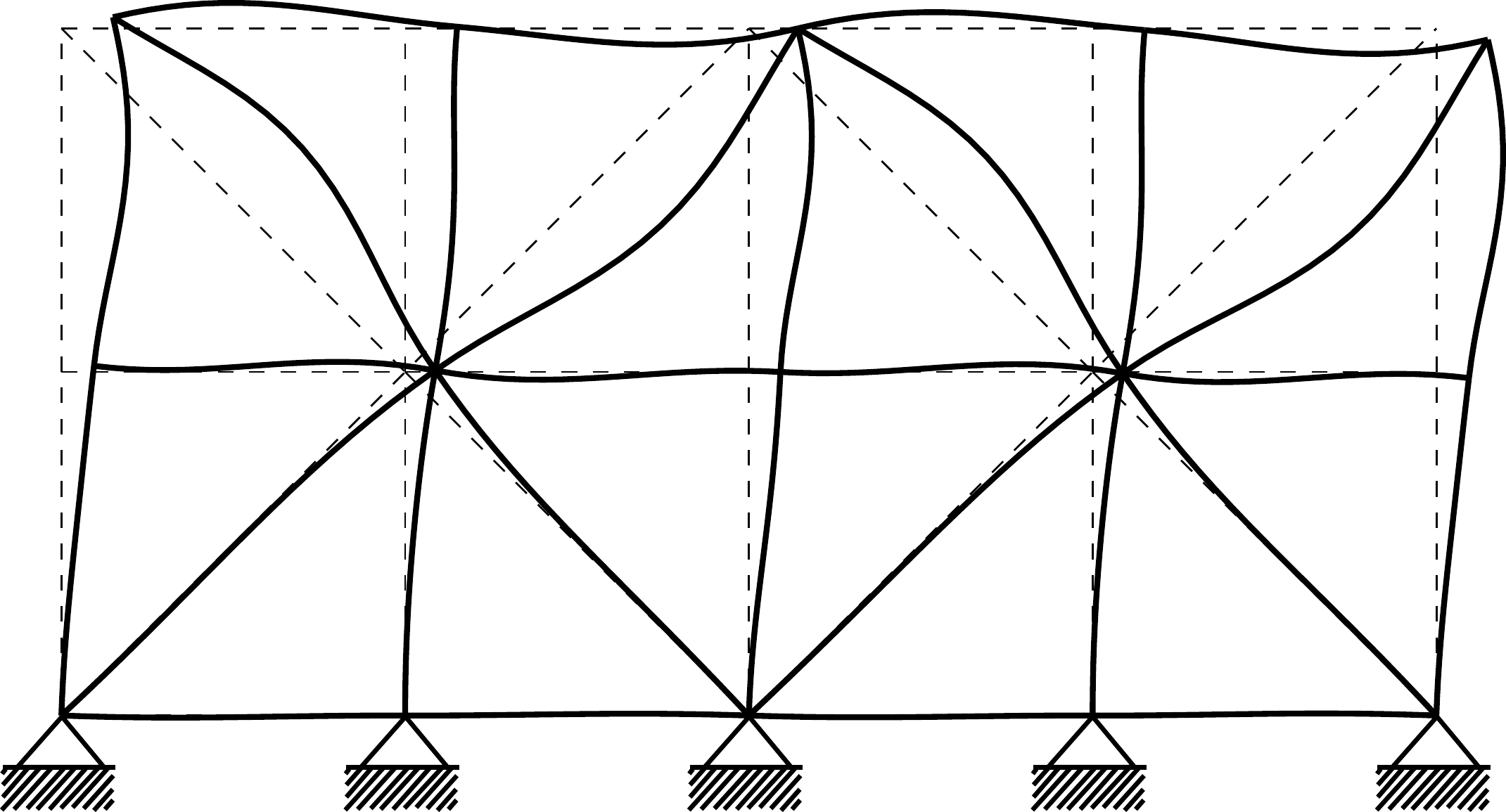}}
\caption{\label{fig:mode_1_flex}$1^{\mathrm{st}}$ mode}
\end{subfigure}
\begin{subfigure}[pos=t]{.45\linewidth}
\includegraphics[scale=0.32]{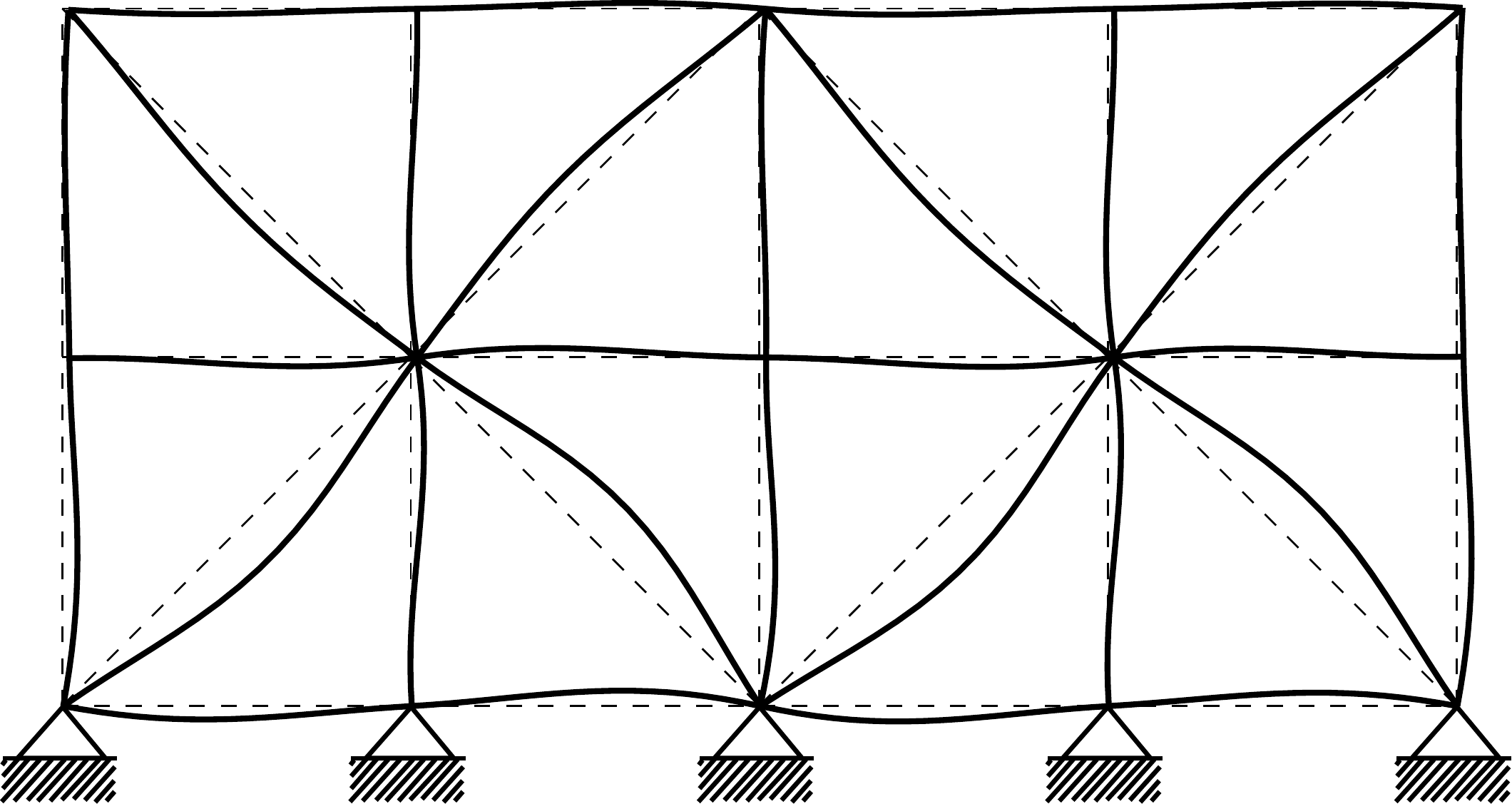}
\caption{\label{fig:mode_2_flex}$2^{\mathrm{nd}}$ mode}
\end{subfigure}\\\vspace{1cm}
\begin{subfigure}[pos=t]{.45\linewidth}
\includegraphics[scale=0.32]{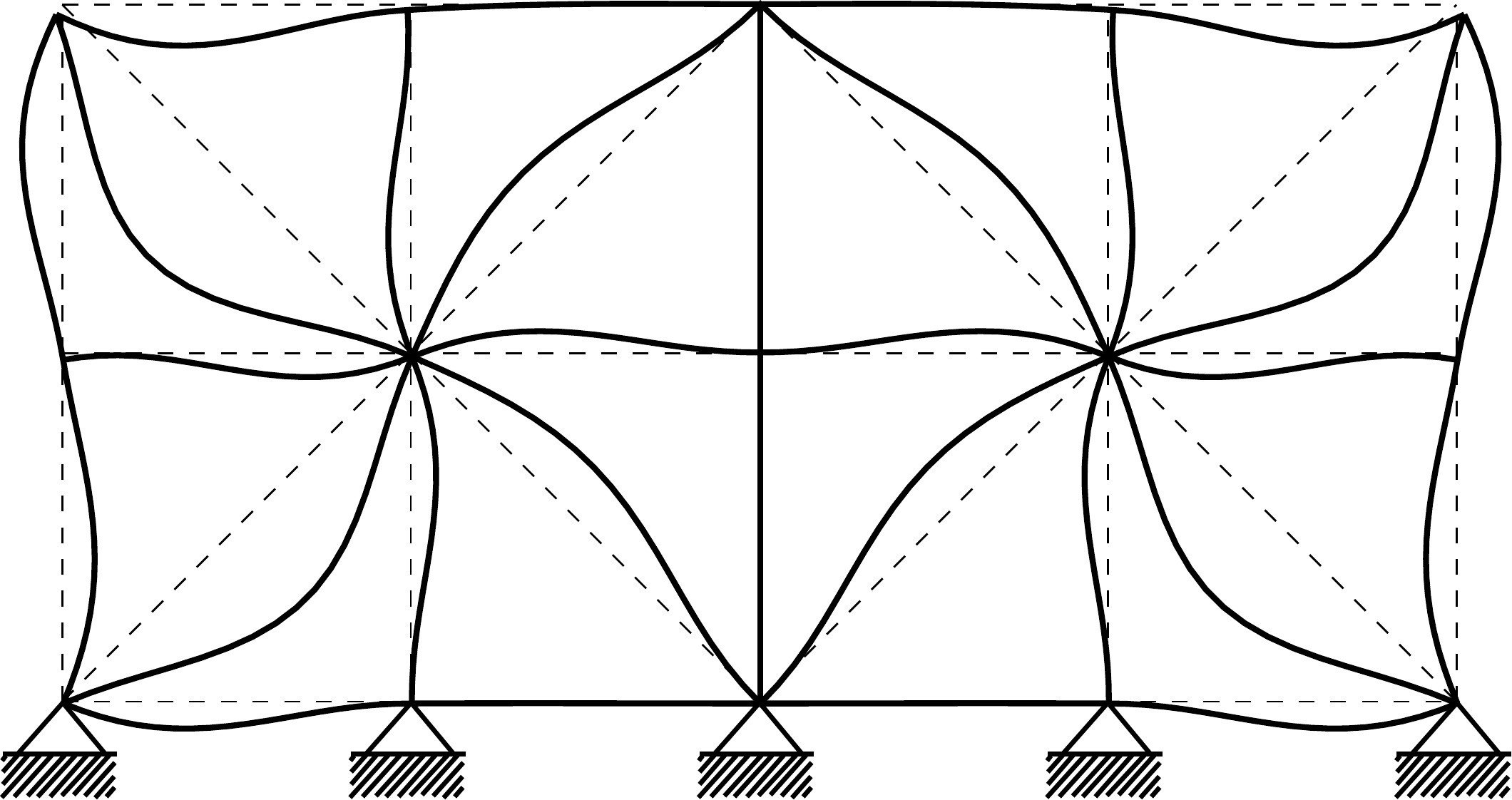}
\caption{\label{fig:mode_3_flex}$3^{\mathrm{rd}}$ mode}
\end{subfigure}
\begin{subfigure}[pos=t]{.45\linewidth}
\includegraphics[scale=0.32]{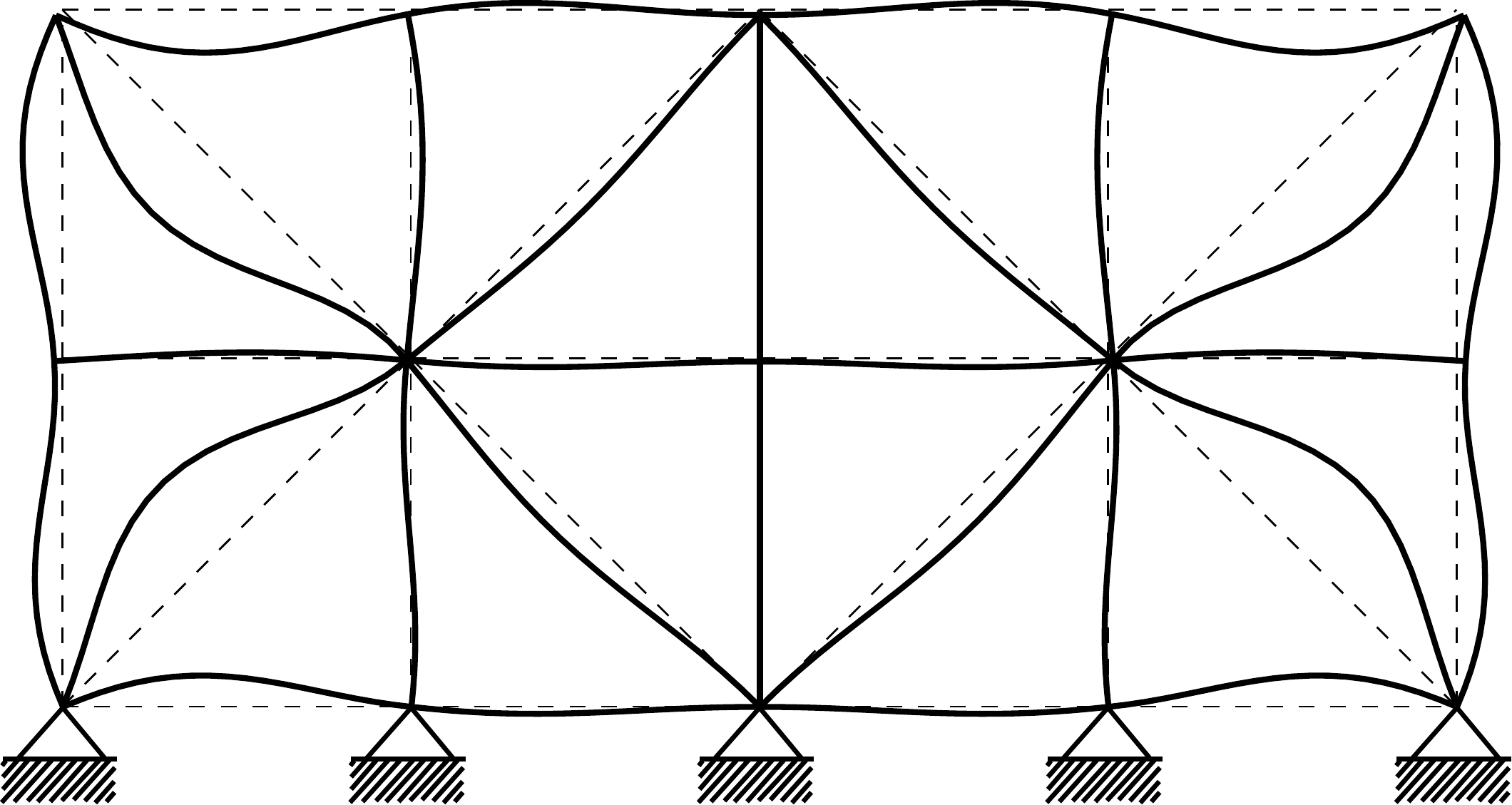}
\caption{\label{fig:mode_5_flex}$5^{\mathrm{th}}$ mode}
\end{subfigure}
\caption{\label{fig:vibration_mode_flexional}Mode shapes for frame in Fig.~\ref{fig:frame}b, for $\lambda = 0.10$.}
\end{figure}

\begin{figure}
\centering
\begin{subfigure}[pos=h]{\linewidth}
\centering
\includegraphics[scale=0.8]{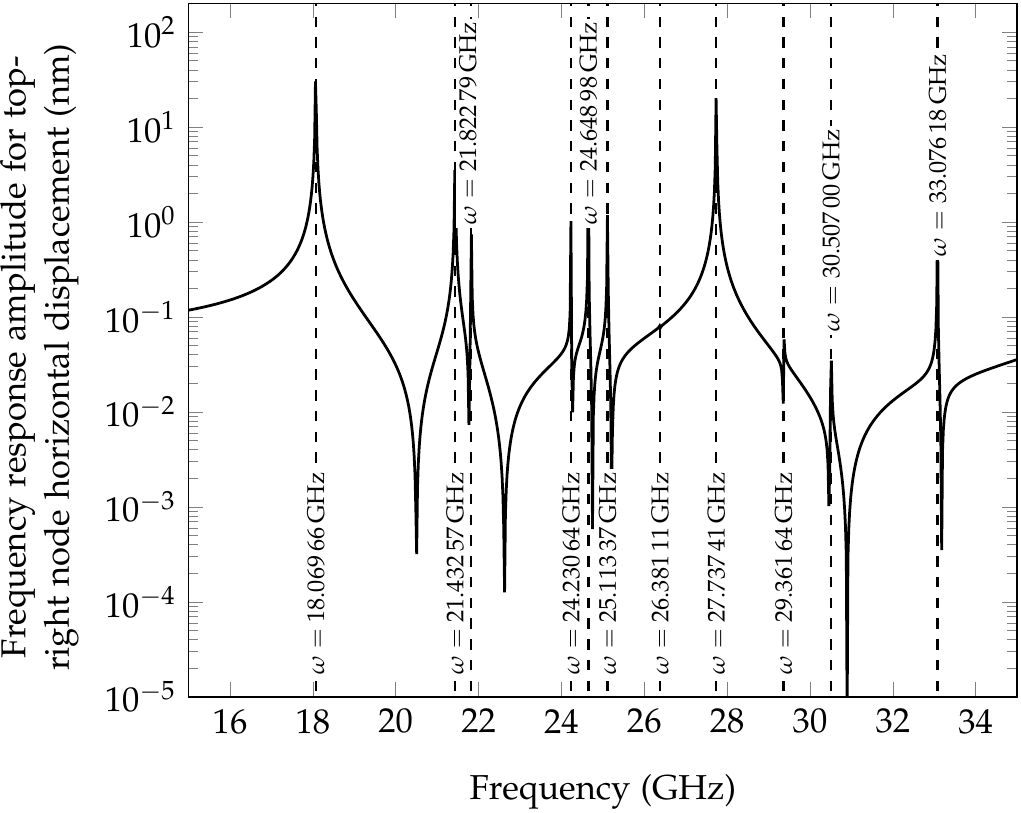}
\end{subfigure}\\\vspace{0.5cm}
\begin{subfigure}[pos=h]{\linewidth}
\centering
\includegraphics[scale=0.8]{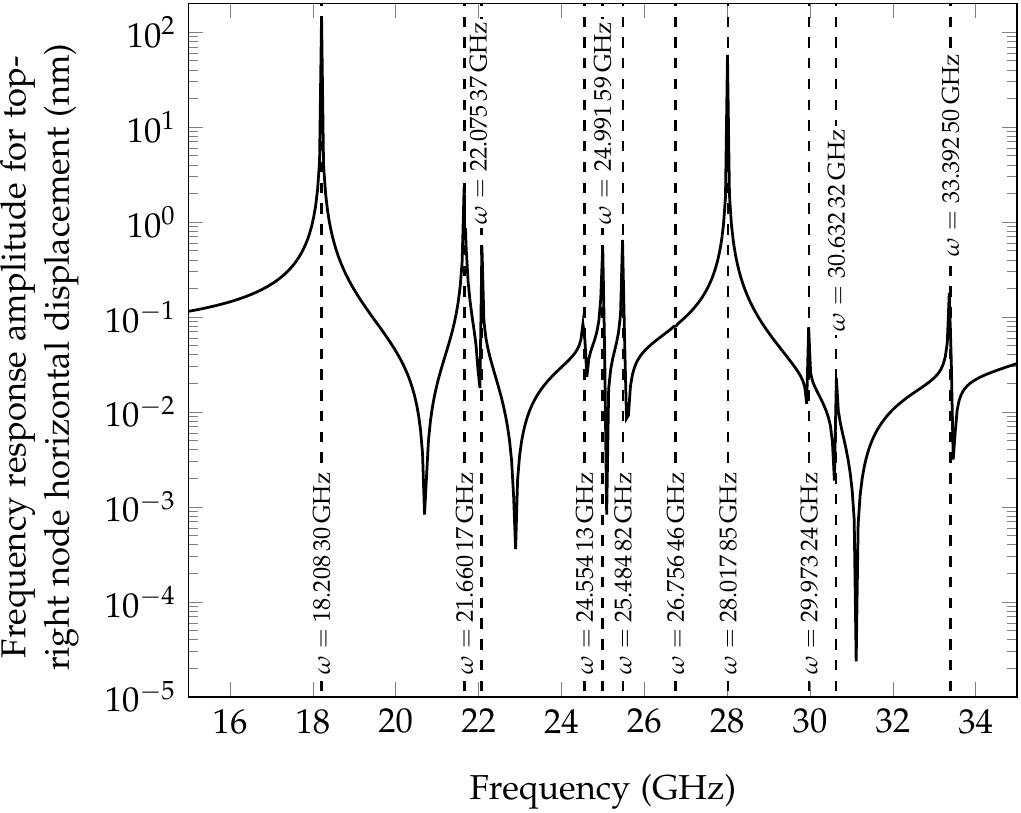}
\end{subfigure}\\\vspace{0.5cm}
\begin{subfigure}[pos=h]{\linewidth}
\centering
\includegraphics[scale=0.8]{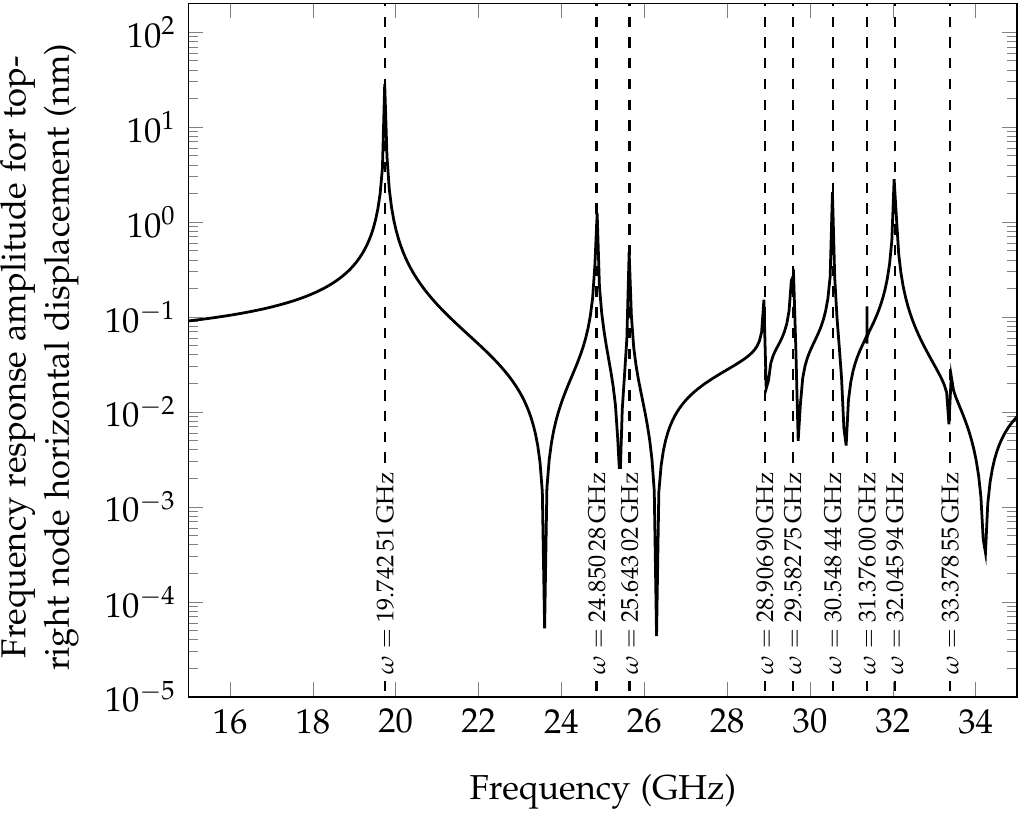}
\end{subfigure}\\
\caption{\label{fig:FRF}Frequency response function for horizontal displacement of top-right node of frame in Fig.~\ref{fig:frame}b, for various internal lengths: (a) $\lambda = 0$, (b) $\lambda = 0.01$, (c) $\lambda = 0.10$.}
\end{figure}

\begin{table}[pos=ht]
\caption{Natural frequencies, $J_{0}$ and $s[\vect{D}_{G}]$ for frame in Fig.~\ref{fig:frame}b.}\label{tbl:beams}
\begin{tabular}{cccccccccc}
\toprule[1pt]
$\lambda $&\multicolumn{3}{c}{0}&\multicolumn{3}{c}{0.01}&\multicolumn{3}{c}{0.10}\\
\midrule[0.5pt]
&$\omega~(\SI{}{\GHz})$&$J_0$&$s[\vect{D}_{G}]$&$\omega~(\SI{}{\GHz})$&$J_0$&$s[\vect{D}_{G}]$&$\omega~(\SI{}{\GHz})$&$J_0$&$s[\vect{D}_{G}]$\\
\cmidrule[0.5pt](lr{0.125em}){2-4}\cmidrule[0.5pt](lr{0.125em}){5-7}\cmidrule[0.5pt](lr{0.125em}){8-10}
 & 18.06966 & 0 & 1 & 18.20830 & 0 & 1 & 19.74251 & 0 & 1 \\ 
 & 21.43257 & 0 & 2 & 21.66017 & 0 & 2 & 24.85028 & 0 & 2 \\ 
 & 21.82279 & 0 & 3 & 22.07537 & 0 & 3 & 25.64302 & 0 & 3 \\ 
 & 24.23064 & 0 & 4 & 24.55413 & 0 & 4 & 28.90690 & 0 & 4 \\ 
 & 24.64898 & 0 & 5 & 24.99159 & 0 & 5 & 29.58275 & 0 & 5 \\ 
 & 25.11337 & 0 & 6 & 25.48482 & 0 & 6 & 30.54844 & 0 & 6 \\ 
 & 26.38111 & 0 & 7 & 26.75646 & 0 & 7 & 31.37600 & 0 & 7 \\ 
 & 27.73741 & 0 & 8 & 28.01785 & 0 & 8 & 32.04594 & 0 & 8 \\ 
 & 29.36164 & 0 & 9 & 29.97324 & 0 & 9 & 33.37855 & 0 & 9 \\ 
 & 30.50700 & 8 & 2 & 30.63232 & 8 & 2 & 37.90881 & 0 & 10 \\ 
 & 33.07618 & 8 & 3 & 33.39250 & 8 & 3 & 39.42049 & 0 & 11 \\ 
 & 36.73234 & 8 & 4 & 37.08584 & 8 & 4 & 42.08654 & 8 & 4 \\ 
 & 36.90634 & 8 & 5 & 37.29529 & 8 & 5 & 42.64020 & 8 & 5 \\ 
 & 37.22135 & 8 & 6 & 37.67109 & 8 & 6 & 44.91988 & 8 & 6 \\ 
 & 39.65386 & 8 & 7 & 40.09235 & 8 & 7 & 46.64787 & 8 & 7 \\ 
 & 40.83422 & 8 & 8 & 41.28685 & 8 & 8 & 48.15536 & 8 & 8 \\ 
 & 41.20524 & 8 & 9 & 41.74052 & 8 & 9 & 49.12239 & 8 & 9 \\ 
 & 42.17990 & 8 & 10 & 42.71746 & 8 & 10 & 49.56440 & 8 & 10 \\ 
 & 44.39416 & 8 & 11 & 44.97509 & 8 & 11 & 52.63650 & 8 & 11 \\ 
 & 45.49144 & 8 & 12 & 46.10350 & 8 & 12 & 54.30830 & 8 & 12 \\
\cmidrule[1pt]{2-10}
\end{tabular}
\end{table}
Mode shapes associated with some of the natural frequencies in Table~\ref{tbl:beams}, reported in Fig.~\ref{fig:vibration_mode_flexional}, exhibit symmetry and anti-symmetry as expected in vibrating structures. 

\clearpage
\section{Closing remarks}
A novel approach to the dynamics of small-size frames has been proposed, resorting to the analysis presented in \citep{ROMANO201714} within special framework of straight beams. 
On adopting a stress-driven nonlocal formulation to account for size effects, the exact dynamic stiffness matrix of a two-node nonlocal element has been analytically derived in a closed form, which can readily be used to construct, by a standard finite-element assembly procedure, the global dynamic stiffness matrix of complex small-size frames. The Wittrick-Williams algorithm has been applied to calculate all natural frequencies and related modes.
The formulation has been presented for 2D structures, but can be generalized to 3D ones. 

The proposed approach provides, to the best of authors' knowledge, a first example of two-node nonlocal element, whose dynamics is treated exactly using the dynamic-stiffness matrix approach. That is, every member of the frame can be modelled by a single, exact element without any internal mesh. The stress-driven approach offers a consistent nonlocal description of size effects that does overcome theoretical flaws of alternative nonlocal formulations. Finally, using the Wittrick-Williams technique guarantees that all natural frequencies can be calculated without missing anyone and including multiple ones. It is believed that the proposed methodology provides an effective and robust tool to assess scale effects in nano-frames, within the general framework of nonlocal mechanics.

\appendix

\section*{Appendix A}
This Appendix reports closed analytical expressions of functions $f_k$ and $g_k$ involved in Eq.~\eqref{eqn:u} and Eq.~\eqref{eqn:v} of the main text.
First, notice that the solution of the $4^{\mathrm{th}}$ order differential equation \eqref{eq16GA} takes the general expression:
\begin{equation}\label{eqn:u_full}
U(\xi) = \sum_{k=1}^{2} c_{a,k} \mathrm{e}^{\sqrt{r_k} \xi} + c_{a,k+2} \mathrm{e}^{-\sqrt{r_k} \xi} 
\end{equation}
where $c_{a,k}$ (for $k=1,...,4$) are integration constants and $\pm \sqrt{r_k}$ (for $k=1,...,2$) are the roots of the characteristic polynomial of Eq.~\eqref{eq16GA}, obtained in the following form on assuming the solution $U=\mathrm{e}^{y\xi}$ and setting $y^2=r$:
\begin{equation}\label{eqn:poly_axial}
-\lambda^2 r^2 + r + \overline{\omega}_a^2 = 0 
\end{equation}
Enforcing the constitutive BCs~\eqref{eqn:dimensionless_gen_strain}, \eqref{eqn:u_full} reverts to Eq.~\eqref{eqn:u} where functions $f_k$ are:
\begin{equation}\label{eqn:f}
\begin{aligned}
&f_{1}(\xi) = \mathcal{A}_1 \mathrm{e}^{-r_1 \xi}+\mathcal{A}_2 \mathrm{e}^{-r_2 \xi}+\mathrm{e}^{r_1 \xi}\\
&f_{2}(\xi) = \mathcal{A}_3 \mathrm{e}^{-r_1 \xi}+\mathcal{A}_4 \mathrm{e}^{-r_2 \xi}+\mathrm{e}^{r_2 \xi}
\end{aligned}
\end{equation}
In Eq.~\eqref{eqn:f}, symbols  $A_{k}$ (for $k=1,...,4$) and $r_k$ (for $k=1,2$)
\begin{equation}
\begin{aligned}
&\mathcal{A}_1=\mathcal{O}_1^{-1} \mathrm{e}^{r_1} \left(\lambda ^2 \mathrm{e}^{r_1+r_2} r_1 r_2-\lambda ^2 r_1 r_2+\lambda  \mathrm{e}^{r_1+r_2} r_1+\lambda  r_1+\lambda  \mathrm{e}^{r_1+r_2} r_2+\lambda  r_2+\mathrm{e}^{r_1+r_2}-1\right)\\
&\mathcal{A}_2=-(r_2 \mathcal{O}_1 )^{-1} \mathrm{e}^{r_2} r_1 \left(\lambda ^2 \mathrm{e}^{2 r_1} r_1^2-\lambda ^2 r_1^2+2 \lambda  \mathrm{e}^{2 r_1} r_1+2 \lambda  r_1+\mathrm{e}^{2 r_1}-1\right)\\
&\mathcal{A}_3=( r_1 \mathcal{O}_1)^{-1} \mathrm{e}^{r_1} r_2 \left(\lambda ^2 \mathrm{e}^{2 r_2} r_2^2-\lambda ^2 r_2^2+2 \lambda  \mathrm{e}^{2 r_2} r_2+2 \lambda  r_2+\mathrm{e}^{2 r_2}-1\right)\\ 
&\mathcal{A}_4=-\mathcal{O}_1^{-1} \mathrm{e}^{r_2} \left(\lambda ^2 \mathrm{e}^{r_1+r_2} r_1 r_2-\lambda ^2 r_1 r_2+\lambda  \mathrm{e}^{r_1+r_2} r_1+\lambda  r_1+\lambda  \mathrm{e}^{r_1+r_2} r_2+\lambda  r_2+\mathrm{e}^{r_1+r_2}-1\right)
\end{aligned}
\end{equation}
with
\begin{equation}
\mathcal{O}_1=\lambda ^2 \mathrm{e}^{p_1} p_1 p_2-\lambda ^2 \mathrm{e}^{p_2} p_1 p_2-\lambda  \mathrm{e}^{p_1} p_1-\lambda  \mathrm{e}^{p_2} p_1+\lambda  \mathrm{e}^{p_1} p_2+\lambda  \mathrm{e}^{p_2} p_2-\mathrm{e}^{p_1}+\mathrm{e}^{p_2}
\end{equation}

Further, recognize that the solution of the $6^{\mathrm{th}}$ order differential equation \eqref{eq17GA} takes the general expression (e.g., see \cite{pinnola2020random}):
\begin{equation}\label{eqn:v_full}
V(\xi) = \sum_{k=1}^{3} c_{b,k} \mathrm{e}^{\sqrt{p_k} \xi} + c_{b,k+3} \mathrm{e}^{-\sqrt{p_k} \xi}  
\end{equation}
where $c_{b,k}$ (for $k=1,...,6$) are integration constants and $\pm \sqrt{p_k}$ (for $k=1,...,3$) are the roots of the characteristic polynomial of Eq.~\eqref{eq17GA}, obtained in the following form on assuming the solution $V= \mathrm{e}^{y\xi}$ and setting $y^2=p$:
\begin{equation}\label{eqn:poly_flex}
\lambda^2 p^3 - p^2 + \overline{\omega}_b^4 = 0 
\end{equation}
Enforcing the constitutive BCs~\eqref{eqn:dimensionless_curvature}, \eqref{eqn:v_full} becomes Eq.~\eqref{eqn:v} where functions $g_k$ are:
\begin{equation}\label{eqn:g}
\begin{aligned}
&g_{1}(\xi) =\mathrm{e}^{p_1 \xi}+ \mathcal{B}_{1}\mathrm{e}^{-p_2 \xi} + \mathcal{B}_{2}\mathrm{e}^{-p_3 \xi}\\
&g_{2}(\xi) =\mathrm{e}^{p_2 \xi}+ \mathcal{B}_{3}\mathrm{e}^{-p_2 \xi} + \mathcal{B}_{4}\mathrm{e}^{-p_3 \xi}\\
&g_{3}(\xi) =\mathrm{e}^{p_3 \xi}+ \mathcal{B}_{5}\mathrm{e}^{-p_2 \xi} + \mathcal{B}_{6}\mathrm{e}^{-p_3 \xi}\\
&g_{4}(\xi) =\mathrm{e}^{-p_1 \xi}+ \mathcal{B}_{7}\mathrm{e}^{-p_2 \xi} + \mathcal{B}_{8}\mathrm{e}^{-p_3 \xi}\\
\end{aligned}
\end{equation}

In Eq.~\eqref{eqn:g}, symbols $\mathcal{B}_{k}$ (for $k =1,...,8$) and $p_{k}$ (for $k=1,...,3$) denote:
\begin{equation}
\begin{aligned}
&\mathcal{B}_{1}=-(p_2^2 \mathcal{O}_2)^{-1} \mathrm{e}^{p_2} p_1^2 \left(\mathrm{e}^{p_1+p_3} p_1 p_3 \lambda ^2-p_1 p_3 \lambda ^2+\mathrm{e}^{p_1+p_3} p_1 \lambda +p_1 \lambda +\mathrm{e}^{p_1+p_3} p_3 \lambda +p_3 \lambda +\mathrm{e}^{p_1+p_3}-1\right)\\ 
&\mathcal{B}_{2}=(p_3^2 \mathcal{O}_2)^{-1} \mathrm{e}^{p_3} p_1^2 \left(\mathrm{e}^{p_1+p_2} p_1 p_2 \lambda ^2-p_1 p_2 \lambda ^2+\mathrm{e}^{p_1+p_2} p_1 \lambda +p_1 \lambda +\mathrm{e}^{p_1+p_2} p_2 \lambda +p_2 \lambda +\mathrm{e}^{p_1+p_2}-1\right)\\
&\mathcal{B}_{3}=-\mathcal{O}_2^{-1} \mathrm{e}^{p_2} \left(\mathrm{e}^{p_2+p_3} p_2 p_3 \lambda ^2-p_2 p_3 \lambda ^2+\mathrm{e}^{p_2+p_3} p_2 \lambda +p_2 \lambda +\mathrm{e}^{p_2+p_3} p_3 \lambda +p_3 \lambda +\mathrm{e}^{p_2+p_3}-1\right)\\
&\mathcal{B}_{4}=(p_3^2 \mathcal{O}_2)^{-1} \mathrm{e}^{p_3} p_2^2 \left(\mathrm{e}^{2 p_2} \lambda ^2 p_2^2-\lambda ^2 p_2^2+2 \mathrm{e}^{2 p_2} \lambda  p_2+2 \lambda  p_2+\mathrm{e}^{2 p_2}-1\right)\\
&\mathcal{B}_{5}=-(p_2^2 \mathcal{O}_2)^{-1} \mathrm{e}^{p_2} p_3^2 \left(\mathrm{e}^{2 p_3} \lambda ^2 p_3^2-\lambda ^2 p_3^2+2 \mathrm{e}^{2 p_3} \lambda  p_3+2 \lambda  p_3+\mathrm{e}^{2 p_3}-1\right)\\
&\mathcal{B}_{6}=\mathcal{O}_2^{-1} \mathrm{e}^{p_3} \left(\mathrm{e}^{p_2+p_3} p_2 p_3 \lambda ^2-p_2 p_3 \lambda ^2+\mathrm{e}^{p_2+p_3} p_2 \lambda +p_2 \lambda +\mathrm{e}^{p_2+p_3} p_3 \lambda +p_3 \lambda +\mathrm{e}^{p_2+p_3}-1\right)\\
&\mathcal{B}_{7}=-(p_2^2 \mathcal{O}_2)^{-1} \mathrm{e}^{p_2-p_1} p_1^2 \left(\mathrm{e}^{p_1} p_1 p_3 \lambda ^2-\mathrm{e}^{p_3} p_1 p_3 \lambda ^2-\mathrm{e}^{p_1} p_1 \lambda -\mathrm{e}^{p_3} p_1 \lambda +\mathrm{e}^{p_1} p_3 \lambda +\mathrm{e}^{p_3} p_3 \lambda -\mathrm{e}^{p_1}+\mathrm{e}^{p_3}\right)\\
&\mathcal{B}_{8}=(p_3^2 \mathcal{O}_2)^{-1} \mathrm{e}^{p_3-p_1} p_1^2 \left(\mathrm{e}^{p_1} p_1 p_2 \lambda ^2-\mathrm{e}^{p_2} p_1 p_2 \lambda ^2-\mathrm{e}^{p_1} p_1 \lambda -\mathrm{e}^{p_2} p_1 \lambda +\mathrm{e}^{p_1} p_2 \lambda +\mathrm{e}^{p_2} p_2 \lambda -\mathrm{e}^{p_1}+\mathrm{e}^{p_2}\right)\\
\end{aligned}
\end{equation}
being
\begin{equation}
\mathcal{O}_2=\mathrm{e}^{p_2} p_2 p_3 \lambda ^2-\mathrm{e}^{p_3} p_2 p_3 \lambda ^2-\mathrm{e}^{p_2} p_2 \lambda -\mathrm{e}^{p_3} p_2 \lambda +\mathrm{e}^{p_2} p_3 \lambda +\mathrm{e}^{p_3} p_3 \lambda -\mathrm{e}^{p_2}+\mathrm{e}^{p_3}
\end{equation}

\section*{Appendix B}
This Appendix describes the formulation of the eigenvalue problem for the free vibrations of stress-driven nonlocal beams, according to the Rayleigh-Ritz method \citep{meirovitch1997principles} using Chebysh\"{e}v polynomials \citep{mason2002chebyshev} as trial functions.

For axial vibrations, the Rayleigh's quotient is defined by
\begin{equation}
\label{eqn:Rayleigh_quotient_axial}
\overline{\omega}_{a}^{2} = R(U) = \frac{\int_{0}^{1}\left(\lambda^2 \mtder{4}{U}{\xi}U - \mtder{2}{U}{\xi}\right)\D{\xi}}{\int_{0}^{1}U^{2}\D{\xi}}
\end{equation}
where $U$ is an eigenfunction. Next, integrate by parts the numerator of Eq.~\eqref{eqn:Rayleigh_quotient_axial}, enforce the constitutive BCs \eqref{eqn:dimensionless_gen_strain} along with the static BCs and assume that $U(\xi)$ is written as an expansion of $N$ shifted Chebysh\"{e}v polynomials of the first kind:
\begin{equation}\label{eqn:expansion_axial}
U(\xi) = \vect{c}^{\mathrm{T}}\vecg{\phi}(\xi)
\end{equation}
where $\vect{c} = \begin{bmatrix}c_0&\dots&c_N\end{bmatrix}^{\mathrm{T}}$ is the vector of constants and $\vecg{\phi}(\xi)$ is the vector
\begin{equation}\label{eqn:basis_vector}
\vecg{\phi}(\xi) = \begin{bmatrix}
T_0(-1+2\xi)&\dots&T_{N}(-1+2\xi)
\end{bmatrix}^{\mathrm{T}}
\end{equation}
being $T_j$ the $j^{\mathrm{th}}$ Chebysh\"{e}v polynomial defined by the following recurrence relation \citep{mason2002chebyshev}:
\begin{equation}
T_{j}(x) = 2x T_{j-1}(x) - T_{j-2}(x), \qquad j =2,3,...
\end{equation}
with $T_{0}(x) = 1$ and $T_{1}(x) = x$.
Using Eq.~\eqref{eqn:expansion_axial}, the stationary condition of the Rayleigh's quotient implies that
\begin{equation}\label{eqn:stationarity}
\pder{R(\vect{c})}{\vect{c}} = \vect{0}
\end{equation}
Eq.~\eqref{eqn:stationarity} leads to following generalized eigenproblem:
\begin{equation}
\left(\vect{A}_{a}-\overline{\omega}_{a}^2\vect{B}\right)\vect{c} = \vect{0}
\label{eqn:axial_linear}
\end{equation}
where
\begin{equation}
\vect{A}_{a} = \int_{0}^{1}\left(\lambda^2 \mtder{2}{\vecg{\phi}}{\xi}\otimes  \mtder{2}{\vecg{\phi}}{\xi} + \tder{\vecg{\phi}}{\xi}\otimes  \tder{\vecg{\phi}}{\xi}\right)\D{\xi}+ \lambda \left(\left.\tder{\vecg{\phi}}{\xi}\otimes  \tder{\vecg{\phi}}{\xi}\right\rvert_{\xi=0}+ \left.\tder{\vecg{\phi}}{\xi}\otimes  \tder{\vecg{\phi}}{\xi}\right\rvert_{\xi=1}\right)
\label{eqn:ax_A}
\end{equation}
and
\begin{equation}
\vect{B} = \int_{0}^{1}\vecg{\phi}\otimes\vecg{\phi}\D{\xi}
\label{eqn:ax_B}
\end{equation}
In a similar fashion, the generalized eigenproblem is derived for bending responses. 
In this case, the Rayleigh's quotient is defined by
\begin{equation}
\label{eqn:Rayleigh_quotient_flexural}
\overline{\omega}_{b}^{4} = R(V) = \frac{\int_{0}^{1}\left(-\lambda^2 \mtder{6}{V}{\xi}V + \mtder{5}{V}{\xi}V\right)\D{\xi}}{\int_{0}^{1}V^{2}\D{\xi}}
\end{equation}
with $V$ an eigenfunction. Again, the numerator of Eq.~\eqref{eqn:Rayleigh_quotient_axial} is integrated by part, the constitutive BCs \eqref{eqn:dimensionless_curvature} along with static BCs are enforced and $V(\xi)$ is written as an expansion of $N$ shifted Chebysh\"{e}v polynomials of the first kind:
\begin{equation}\label{eqn:expansion_flexural}
V(\xi) = \vect{c}^{\mathrm{T}}\vecg{\phi}(\xi)
\end{equation}
where $\vect{c} = \begin{bmatrix}c_0&\dots&c_N\end{bmatrix}^{\mathrm{T}}$ is the vector of constants and $\vecg{\phi}(\xi)$ is the vector \eqref{eqn:basis_vector}. Finally, replacing \eqref{eqn:Rayleigh_quotient_flexural} for $R$ in Eq.~\eqref{eqn:stationarity} leads to the following eigenproblem
\begin{equation}
\left(\vect{A}_{b} - \overline{\omega}_{b}^{4}\vect{B}\right)\vect{c} = \vect{0}
\label{eqn:flexural_linear}
\end{equation}
where
\begin{equation}
\vect{A}_{b} = \int_{0}^{1}\left(\lambda^2 \mtder{3}{\vecg{\phi}}{\xi}\otimes  \mtder{3}{\vecg{\phi}}{\xi} + \mtder{2}{\vecg{\phi}}{\xi}\otimes  \mtder{2}{\vecg{\phi}}{\xi}\right)\D{\xi}+ \lambda \left(\left.\mtder{2}{\vecg{\phi}}{\xi}\otimes  \mtder{2}{\vecg{\phi}}{\xi}\right\rvert_{\xi=0}+ \left.\mtder{2}{\vecg{\phi}}{\xi}\otimes  \mtder{2}{\vecg{\phi}}{\xi}\right\rvert_{\xi=1}\right)
\label{eqn:fl_A}
\end{equation}
and $\vect{B}$ is given by Eq.~\eqref{eqn:ax_B}.
It has to be noticed that, in the formulation of both eigenvalue problems \eqref{eqn:axial_linear} and \eqref{eqn:flexural_linear}, kinematic BCs can be suitably enforced on the trial functions using the method of Lagrange multipliers \citep{CANALES2016136}. This leads to the following eigenvalue problems:
\begin{equation}\label{eqn:final_eigenproblem}\tag{\theequation a,b}
\left(
\begin{bmatrix}
\vect{A}_a&\vect{L}_{a}\\
\vect{L}_{a}^{\mathrm{T}}&\vect{0}
\end{bmatrix}
-
\overline{\omega}_a^2
\begin{bmatrix}
\vect{B}&\vect{0}\\
\vect{0}&\vect{0}
\end{bmatrix}
\right)
\begin{bmatrix}
\vect{c}\\
\vecg{\mu}
\end{bmatrix}
=\vect{0}
\qquad
\left(
\begin{bmatrix}
\vect{A}_b&\vect{L}_{b}\\
\vect{L}_{b}^{\mathrm{T}}&\vect{0}
\end{bmatrix}
-
\overline{\omega}_b^4
\begin{bmatrix}
\vect{B}&\vect{0}\\
\vect{0}&\vect{0}
\end{bmatrix}
\right)
\begin{bmatrix}
\vect{c}\\
\vecg{\mu}
\end{bmatrix}
=\vect{0}
\end{equation}
with $\vect{L}_{a}$ and $\vect{L}_b$ involving the trial functions computed at beam ends according to kinematic BCs. For instance, for a cantilever beam:
\begin{equation}\label{eqn:L}
\vect{L}_{a} = \begin{bmatrix}\vecg{\phi}(0) \end{bmatrix}
 \qquad \vect{L}_{b} = \begin{bmatrix}\vecg{\phi}(0)&
\left.\tder{\vecg{\phi}}{\xi}\right\rvert_{\xi=0} \end{bmatrix}\refstepcounter{equation}\tag{\theequation a,b}
\end{equation}

\textbf{Acknowledgment} - Financial supports from 
MIUR in the framework of the Project PRIN 2017 - code 2017J4EAYB Multiscale Innovative Materials and Structures (MIMS) -
University of Naples Federico II Research Unit 
is gratefully acknowledged.

\bibliographystyle{cas-model2-names}

\bibliography{cas-refs}

\end{document}